%% file: grammarly.tex
\documentclass[12pt]{article}

\usepackage{booktabs}
\usepackage{dcolumn}

\usepackage{graphicx}
\usepackage{longtable}
\usepackage{natbib}
\usepackage{rotating}
\usepackage{tabularx}
\usepackage{amsmath}
\usepackage{longtable}
\usepackage{tabularx}
\usepackage{caption}
\usepackage{setspace}
\usepackage[margin=1in]{geometry}
\usepackage{lscape}
\usepackage{epstopdf}

\usepackage{tikz}
\usetikzlibrary{calc,intersections,through,backgrounds}
\usepackage{pgfplots}
\usepackage{amsmath}
\pgfplotsset{compat=1.15}
\usepackage{fouriernc}
\usepackage{hyperref}

	\tikzstyle{mybox} = [%
	draw=red,%
	thick,%
	rectangle,%
	rounded corners,%
	text width=3cm,%
	align=center]

\usepackage{amsthm, amssymb}

\input{parameter_files/parameters.tex}

\newcommand{\starlanguage}{Significance indicators: $p \le 0.10:*$,
  $p \le 0.05:**$ and $p \le .01:***$.}
\newcommand{\starlanguageexpanded}{Significance: $p \le 0.10:\dagger$, $p \le 0.05:*$,
  $p \le 0.01:**$ and $p \le .001:***$.}

\newcommand{\signalView}{``signaling view''}
\newcommand{\signalViewPeriod}{``signaling view.''}
\newcommand{\clarityView}{``clarity view''}
\newcommand{\clarityViewPeriod}{``clarity view.''}

\newcommand{\Grammarly}{Algorithmic Writing Service}

\hypersetup{
  colorlinks = TRUE,
  citecolor=blue,
  linkcolor=red,
  urlcolor=black
}

\begin{document}


\title{Algorithmic Writing Assistance on Jobseekers' Resumes Increases Hires}

\date{\today}

\author{Emma van Inwegen \\ MIT \and Zanele Munyikwa \\ MIT \and John J. Horton \\ MIT \& NBER }

\maketitle

\begin{abstract}
  \noindent There is a strong association between the quality of the writing in a resume for new labor market entrants and whether those entrants are ultimately hired.
  We show that this relationship is, at least partially, causal: a field experiment in an online labor market was conducted with nearly half a million jobseekers in which a treated group received algorithmic writing assistance.
  Treated jobseekers experienced an \hiredTEpercent{}\% increase in the probability of getting hired.
  Contrary to concerns that the assistance is taking away a valuable signal, we find no evidence that employers were less satisfied.
  We present a model in which better writing is not a signal of ability but helps employers ascertain ability, which rationalizes our findings.
\end{abstract}

\onehalfspacing

\section{Introduction}

For most employers, the first exposure to a job candidate is typically a written resume.
The resume contains information about the applicant---education, skills, past employment, and so on---that the employer uses to draw inferences about the applicant's suitability for the job.
Conveying this information is the most important function of the resume.
A better-written resume---without any change in the underlying facts---might make it easier for the employer to draw the correct inferences, which could lead to a greater chance of an interview or job offer.
We call this the \clarityView{} of the role of resume writing quality.
However, a resume might not merely be a conduit for match-relevant information; the resume's writing itself could signal ability.
In particular, the quality of the writing might be informative about the jobseeker's ability and communication skills.
This is another reason better writing could lead to a greater chance of an interview or a job offer.
We call this the \signalView{} of the role of resume writing quality.

In this paper, we explore how resume writing quality affects the hiring process using both observational data and a field experiment.
We focus on distinguishing between the \clarityView{} and \signalViewPeriod{}
Using observational data from a large online labor market, we document a strong positive relationship between writing quality and hiring (and not simply callbacks).
This relationship persists even after controlling for other factors that might otherwise explain the relationship. 
In terms of magnitude, an additional 1 percentage point increase in error rate (number of errors in the resume divided by the number of words in the resume) is associated with a 3\% decrease in the probability of being hired.
However, this is only an association and there are other potential reasons why writing quality could be correlated with hiring even with our controls.
As such, we report the results of a field experiment in which we vary writing quality in the same market.

The typical approach to addressing a question of causality in hiring preferences would be an audit study, where the researcher would make fictitious job applications and observe call-back rates.
However, this method of analysis has a number of downsides, such as deception and wasting employers' time \citep{kessler2019incentivized}. Furthermore, a callback is merely the first step in the hiring funnel, making it an imperfect proxy for who actually gets hired.

We use an alternative approach. We intercept jobseekers at the resume-writing stage and randomly offer some of them---the treatment group---algorithmic writing assistance.
Others---the control group---had the status quo experience of no assistance.
This writing assistance creates random variation in writing quality.
The algorithmic writing assistance was provided by a company we call the Algorithmic Writing Company.
We will discuss in depth what the \Grammarly{} provides, but generally, it makes writing better by identifying common errors and offering the writer suggestions on how to address those errors.

In the experimental data, there is a very strong ``first stage,'' in that those treated had better-written resumes on several quantifiable dimensions.
For example, we find fewer spelling and grammar errors in the resumes of the treated group  of jobseekers.
Positive effects on resume quality were concentrated among the low-end of the distribution in writing quality, as jobseekers with already excellent resumes can benefit little from writing assistance.

After creating a resume, jobseekers engage in search, which may or may not lead to a job.
We observe job search behavior and outcomes for both treated and control jobseekers.
Treated workers did not send out more applications than workers in the control group, nor did they propose higher wages.
This is a convenient result because our interest is in employers' decision-making, even though randomization was at the level of the jobseeker.
If jobseekers had altered their application behavior---perhaps sending more applications because they know they have a stronger case to make---we might wrongly attribute greater job-market success to the resume rather than this endogenous effort.

Our primary outcome of interest is the effects of writing assistance on hiring.
We find that treated jobseekers had a~\hiredTEpercent\% increase in their probability of being hired at all relative to the control group. The 95\% confidence interval on the percentage increase in hiring is  $(3\%, 13\%.)$
They also had \contractsTEpercent{}\% more job offers over the experimental period than those in the control group.
In terms of the matches themselves, treated workers' hourly wages were~\wagesTEpercent{}\% higher than the hourly wages of workers in the control group.
However, it is important to remember this is a conditional result and could simply be due to composition changes in which workers are hired.

In the \signalView{} the treatment removed or at least weakened a credible signal of jobseeker ability.
If this is the case, this should leave employers disappointed. Unique to our setting, we have a measure of employer disappointment, as both sides privately rate each other at the conclusion of the contract.  Although these ratings have been shown to become inflated over time~\citep{filippas2018} and can be distorted when they are public and reciprocal~\citep*{bolton2013engineering}, they are still a useful signal of worker performance.  If employers are disappointed with the performance of the worker, this would likely manifest in lower employer ratings at the conclusion of the contract.
We find no evidence that this is the case.
If anything, the treatment group had slightly higher ratings.
The average rating of employer satisfaction of workers in the control group was~\privateRatingControl{} on a ten-point scale.
The average rating in the treatment group was \privateRatingTreat{} and had a confidence interval of $(8.74, 8.94)$.
A natural question is how much statistical power we would have to detect differences in the marginal hires induced by the treatment.
Under conservative assumptions, we have 80\% power to detect if marginally induced hires were 0.2 standard deviations worse.
Given the~\wagesTEpercent \% higher average wages in the treatment group, if employers were simply tricked into hiring worse workers generally, these higher wages should have made it even more likely to find a negative effect on ratings~\citep{luca2021effect}.

One possible explanation for our results is that employers are simply wrong in regarding resume writing quality as informative about ability.
However, the \clarityView{} can also rationalize our results without making this assumption.
It is helpful to formalize this notion to contrast it with the more typical signaling framing of costly effort and hiring.
To that end, we present a simple model where jobseekers have heterogeneous private information about their productivity but can reveal their type via writing a ``good'' resume.
This is not a signaling model where more productive workers face lower resume-writing costs---any worker, by writing a good resume, will reveal their information, and this cost is assumed to be independent of actual productivity. Our model has heterogeneous ``good'' resume writing costs. We show that writing assistance shifting the cost distribution can generate our findings of more hires, higher wages, and equally satisfied employers.

Our main contribution is to compare the \clarityView{} and \signalView{} for the positive relationship between writing and hiring.
Our main substantive finding is evidence for the \clarityViewPeriod{}
We can do this because we can trace the whole matching process from resume creation all the way to a measure of post-employment satisfaction.
Helping jobseekers have better-looking resumes helped them get hired (consistent with both explanations), but we find no evidence that employers were later disappointed (which is what the \signalView{} explanation would predict).
We also contribute more broadly by showing the importance of text in understanding matching \citep{marinescu2020opening}.
The notion that better writing can help a reader make a better purchase decision is well-supported in the product reviews literature \citep{ghose2010estimating} but is a novel finding in labor markets.
In one related example,~\cite*{hong2021just} shows that workers who directly message prospective employers (politely) are more likely to get hired, but the politeness effect is muted when the workers' messages contain typographic errors.

In addition to the general theoretical interest in understanding hiring decisions, there are practical implications to differentiating between these two views of the resume.
If the \clarityView{} is more important, then any intervention that encourages better writing is likely to be beneficial.
There will likely be little loss in efficiency if parties are better informed.
Even better, as we show, the kind of assistance that improves clarity can be delivered algorithmically.
These interventions are of particular interest because they have zero marginal cost~\citep{horton2017, beloit2018}, making a positive return on investment more likely, a consideration often ignored in the literature \citep{card2010active}.
On the other hand, if the \signalView{} is more important, then providing such writing assistance will mask important information and lead to poor hiring decisions.

Unlike general advice, algorithmic interventions are adaptive.
In our study, the algorithm took what the jobseeker was trying to write as input and gave targeted, specific advice on improvement.
This is likely more immediately useful than more vague recommendations, such as telling jobseekers to ``omit needless words.''
This advice comes in the form of recommendations that are predicted to improve the resume's effectiveness.  
A limitation of our study is that we cannot speak to crowd-out effects \citep{crepon2013labor}, which are relevant to discuss the welfare implications of any labor market intervention.
However, this concern is somewhat secondary to our narrower purpose of understanding how employers make decisions with respect to resumes. Additionally, given that in our setting, new entrants compete with established jobseekers on the platform, we anticipate the crowd-out effect will be small, and perhaps even welcome if at the expense of more established workers, given the obstacles new entrants face~\citep{pallais2010inefficient}.

In addition to exploring an AI technology in a real labor market, we contribute to a large literature on how variation in applicant attributes affects callback rates \citep{moss2012science, bertrand2003emily, kang2016whitened, farber2016determinants}.
While we are not the first to show that writing matters in receiving callbacks from employers \citep{sterkens2021costly, Martin-Lacroux2017}, we are the first to do so on such a massive scale and with natural variation in writing quality\footnote{
While the reason this preference exists is not known, recruiters report, anecdotally, caring about a resume's writing quality~\citep{oreopoulos2011skilled}.}.
Our experiment involves \numberTotal{} jobseekers which is an order of magnitude larger than the next largest experiments.
Another benefit is that we do not need to guess how workers might make mistakes on their resumes, as it is workers and not researchers writing their resumes.
Additionally, unique in this literature, we can follow the induced changes all the way through hiring and even post-employment assessment which allows us to answer our \clarityView{} vs. \signalView{} questions.

The rest of the paper proceeds as follows.
Section \ref{sec:empirical} describes the online labor market which serves as the focal market for this experiment.
Section~\ref{sec:results} reports the experimental results of the treatment effects on writing quality and subsequent labor market outcomes.
In Section~\ref{sec:discussion} we present a simple model that can rationalize our findings.
Section~\ref{sec:conclusion} concludes.

\section{Empirical context and experimental design} \label{sec:empirical}
The setting for this experiment is a large online labor market.
Although these markets are online, with a global audience, and with lower search costs~\citep{goldfarb2019digital}, they are broadly similar to more conventional markets~\citep{agrawal2015digitization}.
Employers post job descriptions, jobseekers apply, and there are interviews followed by hiring and managing. One distinctive feature of online labor markets is that both the employer and the worker provide ratings for each other at the end of a contract.

Because of the many similarities between on and offline labor markets, a growing body of research uses online labor markets as a setting, often through randomized experiments. 
These studies contribute to the theory in longstanding questions about labor markets, such as deepening our understanding of the mechanisms and processes by which employers and workers find jobs.
Online labor markets also allow researchers to broaden the range of questions in which it is possible to make causal estimates \citep{Horton2010, barach2021employers} because platforms store detailed data on things like applications, text, length of time spent working on an application, speed of hire, and much more.

Many studies on online labor markets identify and measure phenomena that are relevant to labor markets both online and offline. 
Like the offline labor market, online labor markets have been shown to have hiring biases~\citep{chan2018hiring}. 
But,~\cite{agrawal2016online} shows that these biases tend to be ameliorated with experience and that general, employers are able to learn as they hire~\citep{kokkodis2022learning}.
And~\cite{stanton2016landing} shows that in an online labor market, agencies (which act as quasi-firms) help workers find jobs and break into the marketplace.

\subsection{Search and matching on the platform}
A would-be employer writes job descriptions, labels the job opening with a category (e.g., ``Graphic Design''), lists required skills, and then posts the job opening to the platform website.
Jobseekers generally learn about job openings via electronic searches.
They submit applications, including a wage bid and a cover letter.
In addition to jobseeker-initiated applications, employers can also use the interface to search worker profiles and invite workers to apply to particular jobs.
The platform uses the jobseeker's history and ratings on the platform to recommend jobseekers to would-be employers~\citep{horton2017}.
Despite platforms making algorithmic recommendations, none are based on the writing quality of their resume.
In terms of selection, \cite{pallais2010inefficient} shows that employers in an online labor market care about workers' reputation and platform experience when hiring.
After jobseekers submit applications, employers screen the applicants, decide whether to give interviews, and then whether to make an offer(s).

\subsection{Experimental intervention at the resume-writing stage of profile creation}
When new jobseekers sign up to work on the platform, their first step is to register and create their profile. 
This profile serves as the resume with which they apply for jobs.
This profile includes a list of skills, education, and work experience outside of the platform, as well as a classification of their primary job category (e.g., ``Graphic Design''), mirroring what employers select when posting a job.
The interface consists of a text box for a profile title and a longer one for a profile description.
Jobseekers either enter their profile information on the spot or they can copy and paste it from somewhere else.

During the experimental period, jobseekers registering for the platform were randomly assigned to an experimental cell.
The experimental sample comprises jobseekers who joined the platform between June 8th and July 14th, 2021.
For treated jobseekers, the text boxes for the profile description are checked by the \Grammarly{}.
Control jobseekers received the status quo experience.
The experiment included \numberTotal{} jobseekers, with\percentTreat{}\% allocated to the treated cell.
Table~\ref{tab:summary_stats} shows that it was well-balanced and the balance of pre-treatment covariates was consistent with a random process.

\subsection{The algorithmic writing assistance}
\begin{figure}
\caption{Example of the \Grammarly{} 's interface showing suggestions on how to improve writing}
         \includegraphics[width=\textwidth]{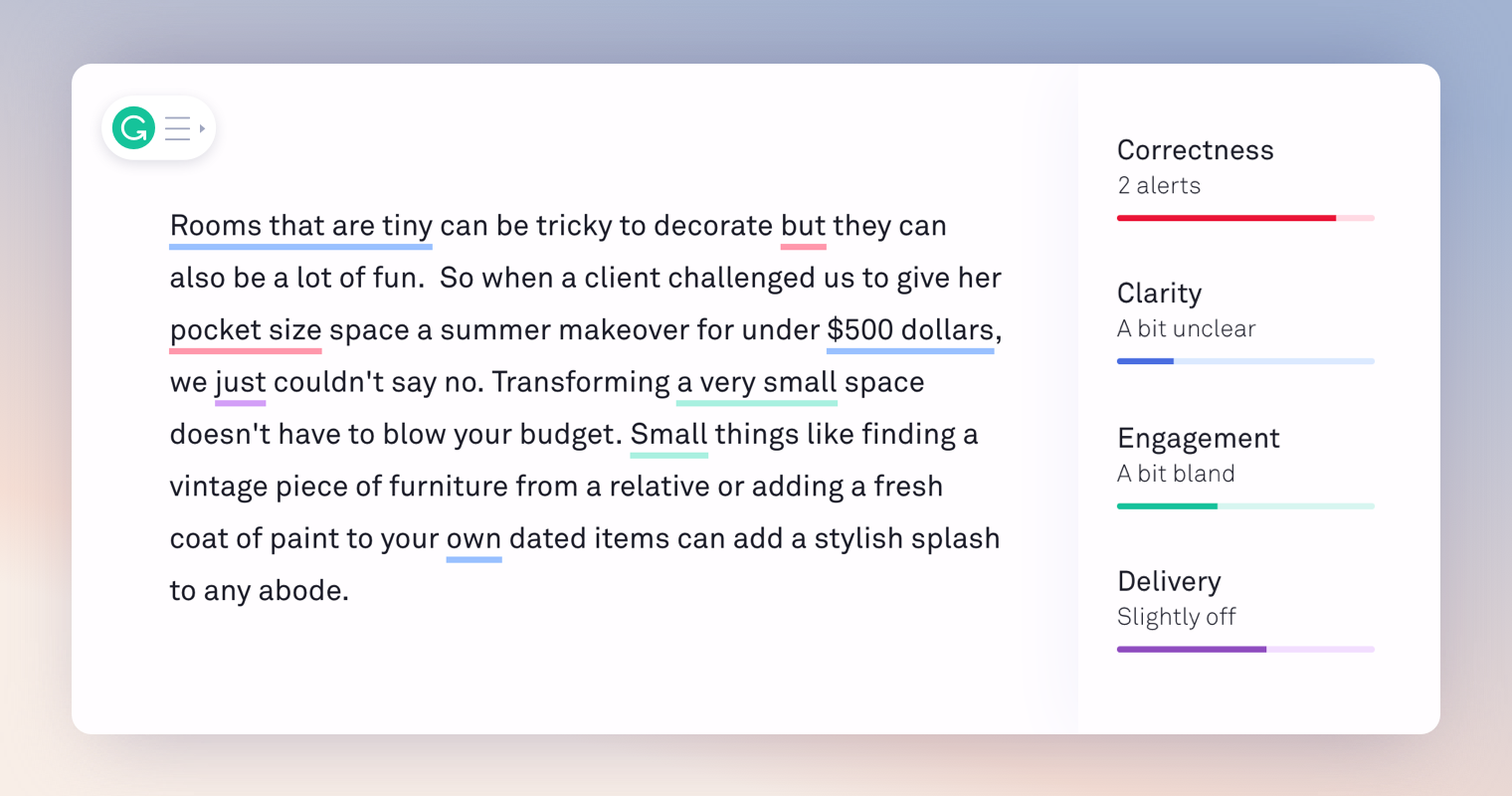}
          \footnotesize Notes: Example of the \Grammarly{} applied to a paragraph of text. To receive the suggestions, users hover their mouse over the underlined word or phrase. For example, if you hover over the first clause ``Rooms that are tiny" underlined in blue,  ``Tiny rooms" will pop up as a suggestion.
         \label{fig:grammerly}
\end{figure}

Words and phrases which are spelled wrong or used incorrectly are underlined by the  \Grammarly{}.
See Figure~\ref{fig:grammerly} for an example of the interface with an example of the text ``marked up'' by the \Grammarly{}.
By hovering a mouse cursor over the underlined word or phrase, the user sees suggestions for fixing spelling and grammar errors.
The \Grammarly{} also gives advice about punctuation, word usage, phrase over-use, and other attributes related to clarity, engagement, tone, and style.

\subsection{Platform profile approval}
When jobseekers finish setting up their profiles, they have to wait to be approved by the platform.
The platform approves jobseekers who have filled out all the necessary information and uploaded an ID and bank details.
The platform can also reject jobseekers at their discretion.
However, platform rejection is somewhat rare.
About 10 percent of profiles are rejected, usually as a part of fraud detection or because the jobseekers leave a completely empty profile.
\percentProfilesSubmitted \% of workers who were allocated into the experiment upon registration complete and submit their profiles.
About~\percentAllProfilesApproved \% of workers who begin registering get all the way through the approval process.

As approval is made following profile creation, this platform step creates a potential problem for interpreting any intervention that changes profile creation.
For example, it could be that better writing just led to a greater probability of platform approval. 
Or, it could have caused jobseekers to be more likely to complete the registration process and submit their profile, both of which could effect hiring.
While unlikely, this is possible, and we do several things to deal with this potential issue.

First, see whether there is any evidence of selection.
We find no evidence that treated jobseekers were more likely to be approved---the estimate is a precise zero.
In Appendix Table~\ref{tab:approved} we show that treated jobseekers are no more likely to submit their profiles and that approval too is unaffected by the treatment.

Second, in our main analysis, we condition on profile approval in our regressions. We also do robustness checks where we report the same analysis not conditioned on profile approval and where we control for profile approval as a covariate.
All our results are robust to these strategies and are described in Section~\ref{sec:robustness}.

Once a jobseeker is approved, they can begin applying for jobs posted on the platform.
Their profile will include their resume and a ``profile hourly wage" which is the wage offer to employers searching for workers.
After they complete their first job on the platform, their profile also shows the worker's actual wages and hours worked on jobs found through the platform.

\begin{singlespace}
\begin{table}[h!]
\caption{Comparison of jobseeker covariates, by treatment assignment \label{tab:summary_stats}}
\begin{footnotesize}
\begin{center}
\begin{tabular}{lllcccc}
\toprule
&   & \multicolumn{1}{c}{\parbox[t][][t]{2cm}{Treatment mean:\\ $\bar{X}_{TRT}$ }}
&     \multicolumn{1}{c}{\parbox[t][][t]{2cm}{Control mean:\\ $\bar{X}_{CTL}$ }}
&     \multicolumn{1}{c}{\parbox[t][][t]{3.3cm}{Difference in means: \\ $\bar{X}_{TRT} - \bar{X}_{CTL}$}}
&     \multicolumn{1}{c}{\parbox[b][][b]{1.2cm}{ p-value }} \\
\\
\hline
  \input{./tables/summary_stats_post.tex} \\

  \multicolumn{3}{l}{\emph{Flow from initial allocation into analysis sample}} \\
   && \emph{Treatment (N)} & \emph{Control (N) } & \emph{Total (N)} && \\
  &Total jobseekers allocated & \numberTreat{} & \numberControl{} & \numberTotal{} & \\
   & $\hookrightarrow$ who submitted their profiles & \numProfilesSubmittedTreat{} & \numProfilesSubmittedControl{} & \numProfilesSubmitted{} & \\
   & \hspace*{0.5cm}$\hookrightarrow$ and were approved by the platform & \numSubProfilesApprovedTreat{} & \numSubProfilesApprovedControl{} & \numSubProfilesApproved  & \\
     & \hspace*{1cm}$\hookrightarrow$ with non-empty resumes & \numProfilesNonEmptyTreat{} & \numProfilesNonEmptyControl{} & \numProfilesNonEmpty{}& \\
       \input{./tables/summary_stats_pre.tex} \\

\bottomrule
 \end{tabular}
 \end{center}
\end{footnotesize}
\singlespace
\begin{footnotesize}
  \emph{Notes:} This table reports means and standard errors of various pre-treatment covariates for the treatment group and the control group.
  The first panel shows the post-allocation outcomes of the full experimental sample i) profile submission, ii) platform approval, iii) length of resume in the number of words, iv) profile hourly wage rate in USD. 
The means of profile hourly rate in treatment and control groups are only for those profiles which report one.
The reported p-values are for two-sided t-tests of the null hypothesis of no difference in means across groups.
The second panel describes the flow of the sample from the allocation to the sample we use for our experimental analysis. 
The complete allocated sample is described in the first line, with each following line defined cumulatively.
The third panel looks at pre-allocation characteristics of the jobseekers in the sample we use for our analysis, allocated jobseekers with non-empty resumes approved by the platform.
  We report the fraction of jobseekers i) from the US, UK, Canada, or Australia, ii) from the US only, iii) specializing in writing jobs, iv) specializing in software jobs, and v) the mean length of their resumes in the number of words.
\end{footnotesize}
\end{table}
\end{singlespace}

\subsection{Description of data used in the analysis}

The dataset we use in the analysis consists of the text of jobseekers' resumes as well as all of their behavior on the platform between the time they registered and August 14th, 2021, one month after allocations ended.
We construct jobseeker level data including the title and text of their profile, the number of applications they send in their first month on the platform, the number of invitations to apply for jobs they receive, the number of interviews they give, and the number of contracts they form with employers.
These workers most often list Design \& Creative, Writing,  Administrative Support, and Software Development as their primary job categories, in order of frequency.

In Table~\ref{tab:summary_stats} we present summary statistics about the jobseekers in the full experimental sample as well as the sample conditioned on platform approval.
16\% of the jobseekers specify that writing jobs are their primary area of work.
Only 14\% of jobseekers are based in the US, and over 80\% are based in a country where English is not the native language.

\subsection{Constructing measures of writing quality}
We do not observe the changes that \Grammarly{} suggested---we simply observe the resumes that result.
As such, we need to construct our own measures of writing quality to determine if the treatment was delivered.

\Grammarly{} gives suggestions to writers about how to improve text along several dimensions.
Perhaps the most straightforward measure of writing quality is spelling.
To see if the treatment impacted spelling errors, we take each worker's profile and check if each word appears in an English language dictionary.
We use the dictionary \texttt{hunspell}, which is based on MySpell dictionaries and is the basis for the spell checker for Google Chrome, Firefox, and Thunderbird.

As many of the resumes are for technical jobs, they often contain industry-specific terms such as ``UX'' or brand names like ``Photoshop.''
To prevent these from being labeled as errors, we augmented the list of words in the dictionary by checking the 1,000 most commonly ``misspelled'' words in our sample and adding non-errors manually.

Spelling is not the only measure of writing quality.
To broaden our measures, we use LanguageTool, an open-source software that finds many errors that a simple spell checker cannot detect, to understand employers care about measures of writing quality other than simply the number of spelling mistakes.
LanguageTool is a rule-based dependency parser that identifies errors (rule violations) and categorizes them.
Some example categories include ``Nonstandard Phrases,'' ``Commonly Confused Words,'' ``Capitalization,'' and ``Typography.''
For example, the nonstandard phrase ``I never have been" would be flagged with a suggestion to replace it with ``I have never been.''
For a more detailed explanation of all of the rule categories, see Appendix Table~\ref{tab:languagetooldescription}. 

\subsection{Spelling errors are associated with lower hiring probabilities in the control group}\label{sec:obs}
\begin{figure}
  \caption{Association between spelling errors and hiring outcomes in the control group}
  \includegraphics[width=\textwidth]{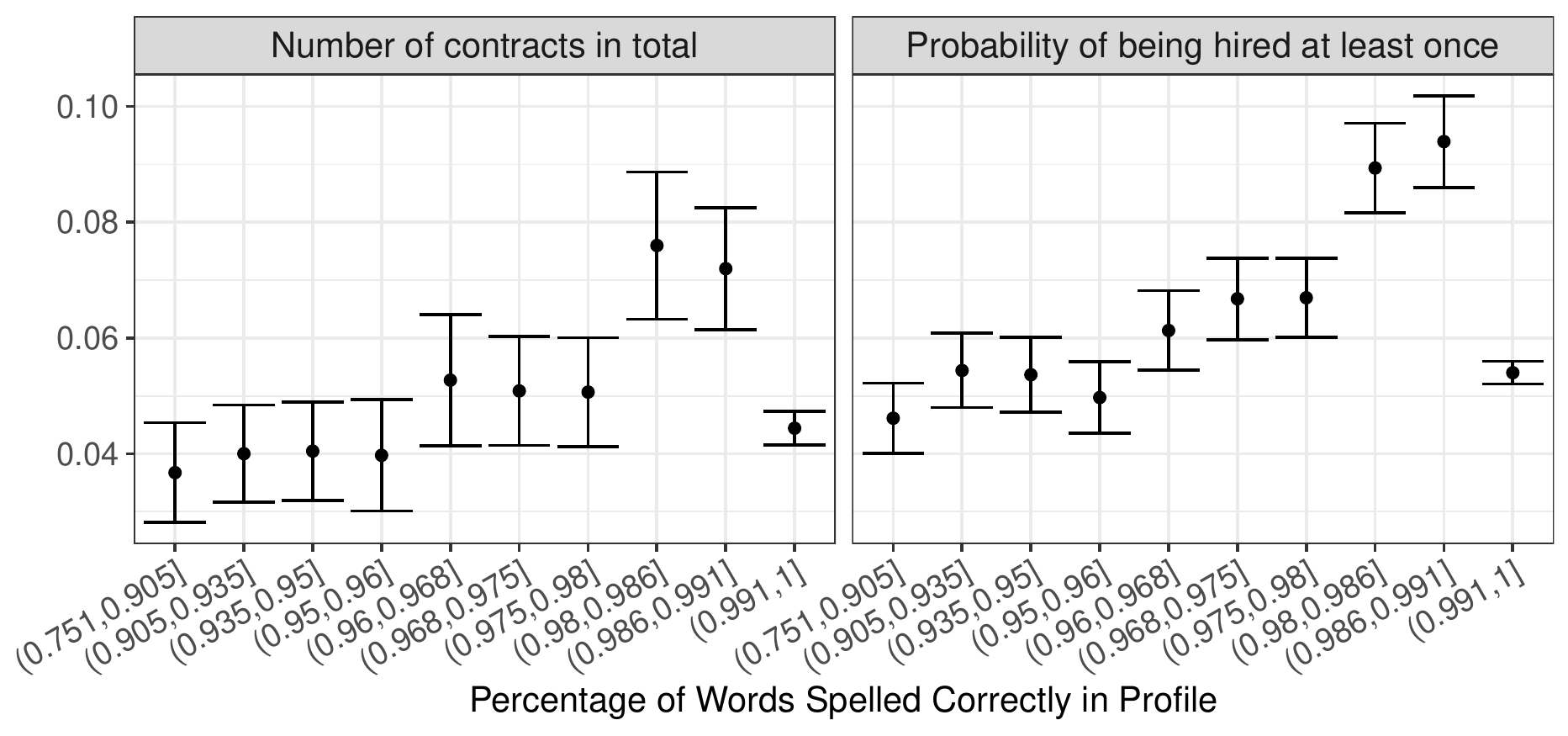}
  \footnotesize \emph{Notes:} These data show the relationship between the percentage of correctly spelled words on a jobseekers' resume with various hiring outcomes.
  A 95\% confidence interval is plotted around each estimate.
 The sample is of all new jobseekers who registered and were approved for the platform between June 8th and July 14th, 2021, and had resumes with more than 10 words. 
 Plotted are jobseekers in the control group, truncated at those who spelled at least 75\% of the words in their resume correctly.
  \label{fig:correlations}
\end{figure}

Before presenting the experimental results, we explore the relationship between resume writing quality and hiring in the control group.
We begin by studying the most unambiguous measure of writing quality: spelling.
In Figure~\ref{fig:correlations} we plot the relationship between hiring outcomes and the percentage of words spelled correctly on the resumes of jobseekers in the control group.
Because the distribution of percent correctly spelled is so left skewed, we truncate the sample to only those who spell at least 75\% of the words in their resumes correctly.
This window includes \percentOver{}\% of jobseekers in the control group.
The x-axis is deciles between 75\% and 100\% of words spelled correctly.

Job seekers with resumes with fewer spelling errors are more likely to be hired.
In the left facet, the y-axis is the number of contracts a jobseeker forms in their first month on the platform.
In the right facet, the y-axis is the probability that a jobseeker is ever hired in their first month on the platform.
A jobseeker with fewer than 90\% of the words in their resume spelled correctly has only a 3\% chance of getting hired, while jobseekers with around 99\% of the words spelled correctly has an 8\% chance of getting hired.
However, as is visible in both facets, resumes with 100\% of words spelled correctly are much \textit{less} likely to receive interest from employers.
This is likely because those resumes tend to be much shorter than the others---the average length of a resume that has zero spelling errors is only \meanLengthNoErrors{} words long.

\subsection{The association between various kinds of writing errors and hiring probabilities}
Moving beyond spelling, in Table~\ref{tab:error_summary_stats} we summarize the occurrence of other types of errors within the control group. In Table~\ref{tab:predict_hiring_languagetool_perword}, we show the correlation between hiring outcomes on each type of language error in the resumes in the control group of the experimental sample.
In these regressions, we control for the jobseekers' profile hourly rate and their job category.
Resumes with more errors in capitalization, grammar, typography, miscellaneous, collocations, possible typo, commonly confused words, and semantics all hired less.
This linear model places some unreasonable assumptions like constant marginal effects on the relationship between various writing errors and hiring.
There may be interactions between these error types.
However, it is still useful to summarize the relationships.
We can see generally negative relationships between writing errors and hiring.

\input{tables/predict_hiring_languagetool_perword}

Interestingly, more style errors \textit{positively} predict hiring.
While initially surprising, style errors are often caused by language being unnecessarily flowery.
Some examples of style errors are ``Moreover, the street is almost entirely residential'' and ``Doing it this way is more easy than the previous method.''
This implies that despite employers' dislike of most writing errors, they forgive or even prefer this kind of flowery language.

\section{Effects of the treatment}\label{sec:results}

We look at two main kinds of experimental results.
First, we examine how the treatment affected the text of resumes.
We are looking to see whether there is a ``first stage.''
Next, we look at market outcomes for those treated workers.
For convenience, we present these treatment effects as percent changes, in Figures~\ref{fig:writing_outcomes} and~\ref{fig:main_outcomes}.
We calculate the standard errors using the delta method.

\subsection{Algorithmic writing assistance improved writing quality}
The first step is to measure the effect \Grammarly{} has on writing in the treatment group.
We start with the fraction of words in the resume spelled incorrectly.
In the control group, resumes are 70 words long on average.
Even the worst spellers spell most of the words correctly, and an average resume has~\meanCorrectRate{}\% of the words spelled correctly.

\begin{figure}
  \caption{Effect of the algorithmic writing assistance on writing quality measures} \label{fig:writing_outcomes}
  \includegraphics[width=\textwidth]{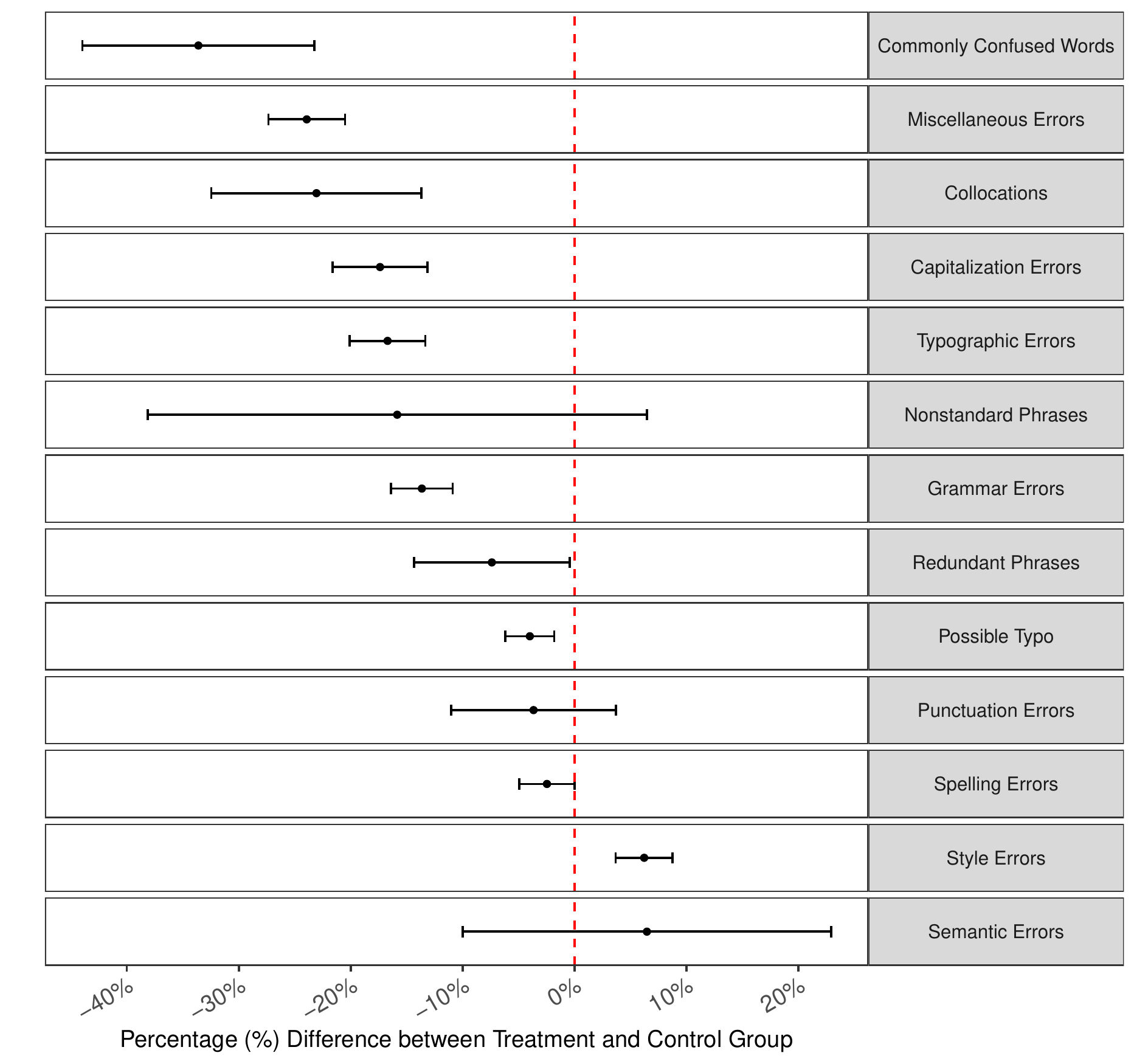}
  \footnotesize \emph{Notes:} This plot shows the effect of the treatment on various writing errors in jobseekers' resumes.
  Point estimates are the percentage change in the dependent variable versus the control group for the treatment groups.
 A 95\% confidence interval based on standard errors calculated using the delta method is plotted around each estimate.
 The experimental sample is of all new jobseekers who registered and were approved for the platform between June 8th and July 14th, 2021, and had non-empty resumes, with $N = $ \numberApprovedNonEmpty .
 Regression details can be found in Appendix Tables~\ref{tab:writing_outcomes_1} and \ref{tab:writing_outcomes_2}.
\end{figure}

To understand the effects of the treatment on other types of writing errors we return to the more fine-grained LanguageTool definitions of writing errors.
In Figure~\ref{fig:writing_outcomes}, we look at the effect of treatment on the number of each type of writing error, normalized by resume length.\footnote{
The treatment had no effect on the length of resumes---see Table~\ref{tab:length} in Appendix~\ref{sec:appendix}.}
Our outcomes of interest are the error rate for each type, so we normalize each type of error to the number of words in the resume.
We calculate the standard errors using the delta method.

We find that jobseekers in the control group had a higher rate of errors of the following types: capitalization, collocations, commonly confused words, grammar, spelling, possible typos, miscellaneous, and typography.
We find larger treatment effects for errors associated with writing clarity than for many others.
For example, the largest magnitudes of differences in error rate were commonly confused words and collocations, where two English words are put together that are not normally found together.
Interestingly, the treatment group had more ``style'' errors.

\subsection{Algorithmic assistance helped the worst writers more}

\begin{figure}
  \caption{Effect of treatment on percentage of words spelled correctly, by deciles}
  \includegraphics[width=\textwidth]{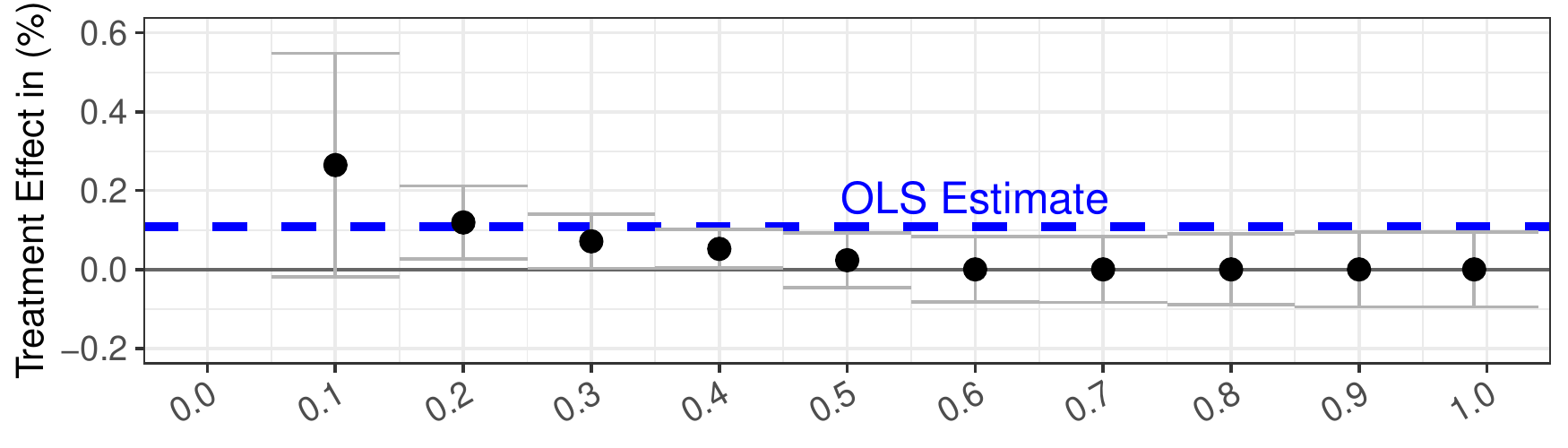}
  \footnotesize Notes: This plot shows the effect of the treatment on the percentage of words spelled correctly in jobseekers' resumes, by deciles. The experimental sample is of all new jobseekers who registered and were approved for the platform between June 8th and July 14th, 2021, and had non-empty resumes, with $N = $ \numberApprovedNonEmpty{}.
  \label{fig:quantiles}
\end{figure}

The treatment was predominantly effective for jobseekers at the bottom of the spelling distribution.
In Figure~\ref{fig:quantiles} we report results from a quantile regression on the effect of the treatment on the percentage of words they spelled correctly.
The effect is concentrated in jobseekers in the bottom half of the spelling distribution.
The treatment effect is largest for jobseekers below the 30\% decile, with effects decreasing at each decile until the median at which point the treatment did not affect spelling.

\begin{figure}
  \caption{Effect of algorithmic writing assistance on hiring outcomes} \label{fig:main_outcomes}
  \includegraphics[width=\textwidth]{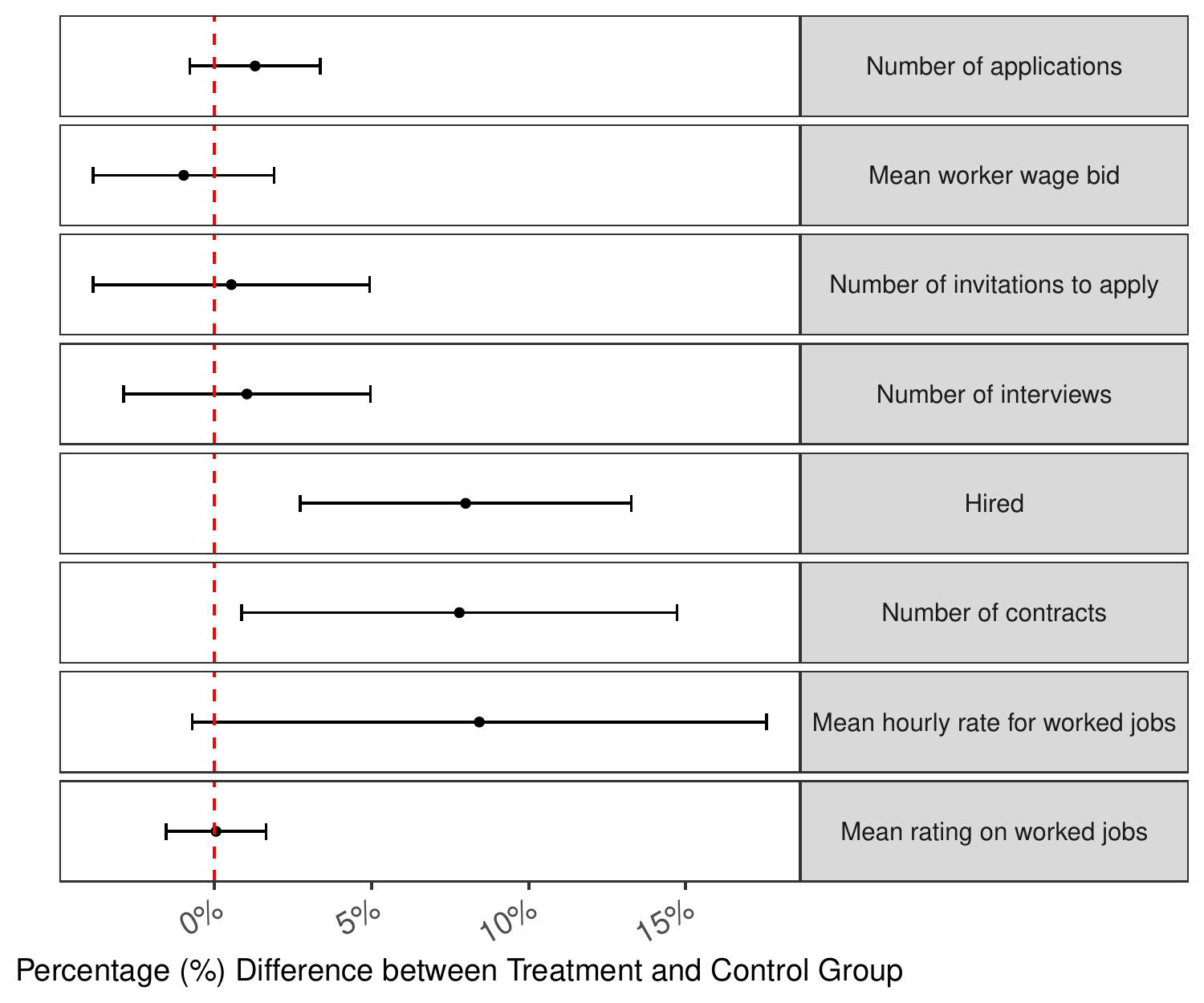}
  \footnotesize \emph{Notes:} This analysis looks at the effect of treatment on hiring outcomes on jobseekers in the experimental sample.
  The x-axis is the difference in the mean outcome between jobseekers in the treated group and the control group.
  A 95\% confidence interval based on standard errors calculated using the delta method is plotted around each estimate.
 The experimental sample is of all new jobseekers who registered and were approved for the platform between June 8th and July 14th, 2021, and had non-empty resumes,  with $N = $ \numberApprovedNonEmpty{}.
 Regression details on the number of applications and wage bid can be found in Table~\ref{tab:applications}.
 Regression details on invitations, interviews, hires, and the number of contracts can be found in Appendix Table~\ref{tab:main_outcomes}.
 Regression details on hourly wages and ratings can be found in Table~\ref{tab:wages_and_rates}.
\end{figure}

\subsection{Heterogeneous treatment effects to spelling}
A natural question is whether effects differed by jobseeker background.
In Table~\ref{tab:treatment} we interact pre-randomization jobseeker attributes with the treatment.
We can see that jobseekers from the US, from English-speaking countries,\footnote{We define whether or not a jobseeker is from a native English-speaking country, as when they login to the platform from USA, UK, Canada, or Australia.} and who are writers all do better in ``levels.''
We find that jobseekers from countries that are not native English speaking experience significantly larger treatment effects to the fraction of words they spell correctly than their anglophone counterparts.

\input{tables/treatment.tex}

\subsection{Treated workers did not change their job search strategy or behavior}
One potential complication in our desire to focus on employer decision-making is that the treatment could have impacted jobseekers search behavior or intensity.
Suppose treated jobseekers changed their behavior, knowing they had higher quality
resumes. In that case, we could not interpret our treatment effect as being driven by employers' having improved perceptions of treated jobseekers. However, we find no evidence that jobseekers changed their search behavior due to the treatment.
In the first facet of Figure~\ref{fig:main_outcomes}, the outcome is the number of applications a jobseeker sends
out over their first 28 days after allocation. We find no effect of the treatment on the number of
applications sent.

In the second facet, the outcome is the mean wage bid proposed by the jobseekers on their applications in their first 28 days on the platform.
Average wage bids in both the treatment and control group were \$\bidsControl{} per hour.
The lack of effects on jobseekers' behaviors makes sense because they were unaware of the treatment.

Table~\ref{tab:applications} show the effects of the treatment on jobseekers application behavior.
In Column (1) we see whether treated jobseekers applied for more jobs than those in the control group over the experimental period and find they did not.
In Column (2) we  find that treated jobseekers do not bid for more hourly jobs than those in the control group.
They also could have bid for higher wages knowing they had better-looking resumes.
In Column (3) we see no evidence of this, where we narrow the sample to only applications to hourly jobs and look at the effect of the treatment on hourly wage bids.

\input{tables/applications.tex}

\subsection{The treatment did not affect employer recruiting}
Employers were able to seek out workers using the platform's search feature to invite jobseekers to apply to their job openings.
In the third facet of Figure~\ref{fig:main_outcomes} from the top, the outcome is the number of invitations to apply for a job that the jobseeker receives in their first month.
We find the effect of the treatment on employer invitations is a precise zero.
In the fourth facet from the top, the outcome is the number of interviews a jobseekers gives over their first month on the platform.
We find that this is also a precise zero.

Although it may seem surprising given the results on hires and contracts, it makes sense given that our experimental sample consists of only new jobseekers to the platform.
New entrants almost never appear in the search results when employers search for jobseekers, given that their rank is determined by their platform history.

\subsection{Treated jobseekers were more likely to be hired}

The treatment raised jobseekers' hiring probability and the number of contracts formed on the platform.
In the fifth facet of Figure~\ref{fig:main_outcomes},  the outcome is a binary indicator for whether or not a jobseeker is ever hired in their first 28 days on the platform.
During the experiment,~\hiredControl\% of jobseekers in the control group worked at least one job on the platform.
Treated jobseekers see an~\hiredTEpercent\% increase in their likelihood of being hired in their first month on the platform.
In Table~\ref{tab:hiring} Column (1) we report these results in levels.

\input{tables/hiring.tex}

Jobseekers in the treated group formed~\contractsTEpercent\% more contracts overall.
In the sixth facet of Figure~\ref{fig:main_outcomes}, the outcome is the number of contracts a jobseeker worked on over their first month.

\subsection{Hourly wages in formed matches were higher}
Treated workers had~\wagesTEpercent\% higher hourly wages than workers in the control group.
In the seventh facet, the outcome is the mean hourly rate workers earned in jobs they worked over their first month on the platform.\footnote{Hourly wage rates for new entrants are not representative of rates on the platform. 
New entrants usually experience rapid wage growth once they gain experience on the platform.}
In the control group, workers on average made \$\wagesControl{} per hour.
In the treatment group, workers made \$\wagesTreatment{} per hour, a significant difference at the~\wagesPvalue{} level.
Since workers did not bid any higher, this result suggests that employers are hiring more productive workers, or that they thought the treated workers were more productive.
If it is the latter, the~\signalView{} would predict that employers would then be disappointed with the workers they hired, which we should be able to observe in worker ratings.

\input{tables/wages_and_rates.tex}

\subsection{Employers gave treated workers slightly better ratings on average}

At the end of every contract, employers rate the workers' quality by reporting a private rating to the platform.
These ratings are not shared with the worker.
In the control group, workers had an average rating of~\privateRatingControl{}.
In the final facet of Figure~\ref{fig:main_outcomes} we show that treated workers who formed any contracts over the experimental period did not have a lower rating than workers in the control group.
We show this result in levels in Table~\ref{tab:wages_and_rates}--- workers in the treated group have an average rating of~\privateRatingTreat{} with a standard error of 0.072.

\subsection{How much power do we have to detect worse contractual outcomes?}
Although the treatment has slightly more positive ratings, a natural question is how much power is available to detect effects.
While we do find a substantial increase in hiring---8\%---these marginal hires are mixed in with a much larger pool of ``inframarginal'' hires that would likely be hired anyway, but for our intervention.
How much worse could those marginal applicants have been and still get our results of slightly higher ratings in the treatment?

\begin{figure}
	\caption{Sampling distribution of the private ratings of marginal hired jobseekers}
         \includegraphics[width=\textwidth]{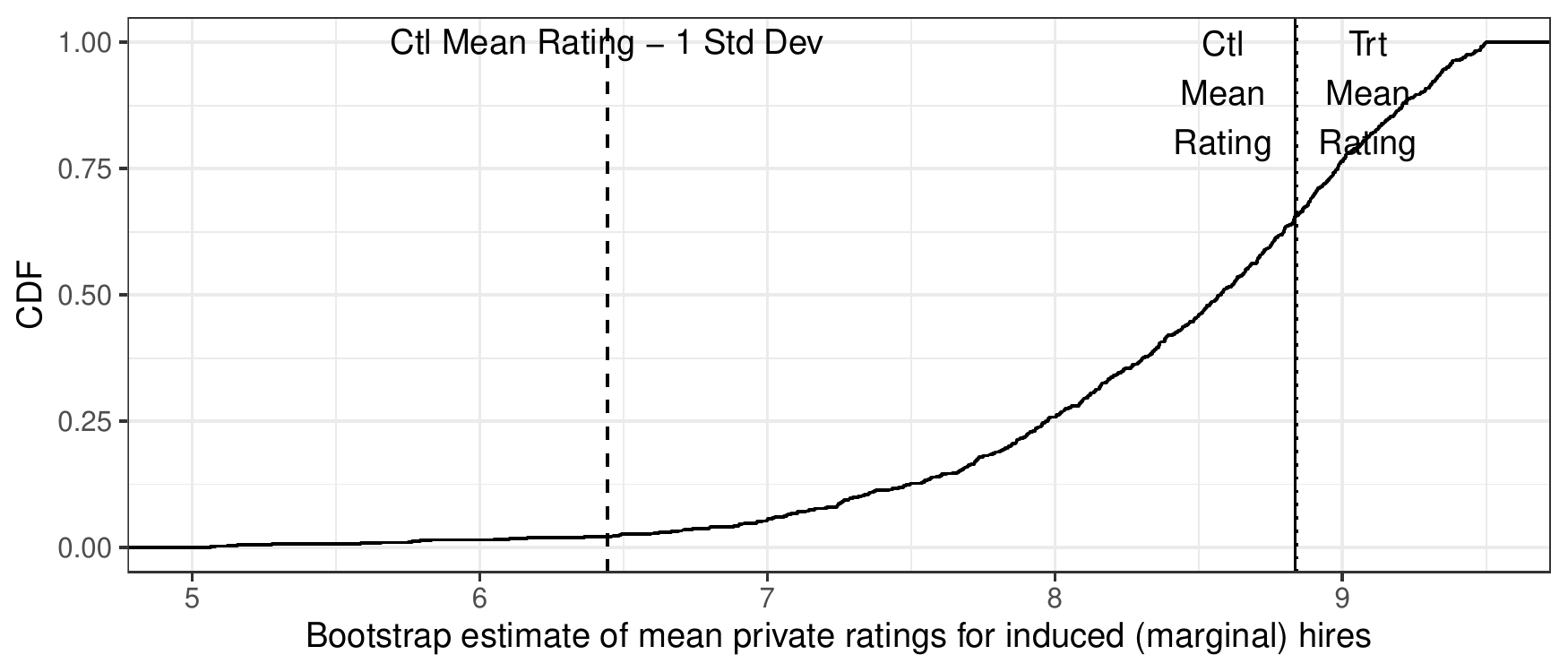}
         \label{fig:bootstrap_sample_distro_of_marginal_rating}
\end{figure}

Let $I$ indicate ``inframarginal'' jobseekers who would have been hired in the treatment or control.
Let $M$ indicate ``marginal'' jobseekers who are only hired in the treatment.
For workers in the control group, the average private rating will be $\bar{r}_C = \bar{r}_I$.
But for the treatment, the mean rating is a mixture of the ratings for the inframarginal and the ratings for the induced, marginal applicants, and so
\begin{align}
  \bar{r}_T = \frac{\bar{r}_I + \tau \bar{r}_M}{1 + \tau}
\end{align}
where $\tau$ is the treatment effect.
We assume no substitution, making our estimates conservative.
The sampling distribution of the mean rating for the marginal group is
\begin{align}
  \bar{r}_M = \frac{\bar{r}_T (1 + \tau) - \bar{r}_C }{\tau}
\end{align}
Our course, $\bar{r}_T$, $\tau$ and $\bar{r}_C$ are all themselves random variables.
Furthermore, they are not necessarily independent.
To compute the sampling distribution of $\bar{r}_M$, we bootstrap sample both the hiring regressions and the private feedback regressions on the experimental sample.\footnote{We define this sample as the workers allocated into the experiment who were approved by the platform and had non-empty resumes. 
From this we bootstrap sample with replacement. 
We run the hiring regressions on this sample and the ratings regressions on the same samples, narrowed to only those workers who were ever hired.}
Because we do not have feedback on workers who are never hired, we use the estimates values to calculate $\bar{r}_M$.
Figure~\ref{fig:bootstrap_sample_distro_of_marginal_rating} shows the sampling distribution of  $\bar{r}_M$.

The treatment and control actual ratings are plotted as solid vertical lines.
As expected given the treatment has a slight positive effect on average ratings, the distribution is centered at these mean values.

The dashed line indicates the control mean rating minus a standard deviation in the private ratings (which is 2.34).
Comparing this value to the distribution of $\bar{r}_M$, this is only about $0.025$ of the density.
In short, it would be quite surprising for us to get the results we have---an 8\% increase in hires and slightly higher (but not significant ratings) if the actual marginal hires were a standard deviation worse.

\subsection{Heterogeneous treatment effects to hiring}

We might have expected the treatment to have differential effects on these subgroups, particularly since  the treatment disproportionately impacted the fraction of words spelled correctly in non-native English speakers' resumes.
In hiring outcomes, we might expect, for example, that native English or US-based jobseekers would benefit less, writers might benefit more---though as we saw earlier, writers already make few errors.
However, for these same jobseekers, the treatment might do less.

We have already shown above in Table~\ref{tab:treatment} that the treatment disproportionately impacted the fraction of words spelled correctly in non-native English speakers' resumes.
If we look downstream to hiring outcomes, in Table~\ref{tab:hiring}, we interact the same groups with the treatment and look at their effect on the probability they were hired.
The point estimates are generally quite imprecise and we lack the power to conclude much.
While non-native English speakers' writing might benefit more from the treatment, it does not translate into more hires relative to native English speakers.

\subsection{Robustness checks}\label{sec:robustness}

In our main analysis narrow the sample to only those jobseekers whose profiles were approved by the platform.
In Appendix Table~\ref{tab:main_outcomes_control} we run a similar regression on the full experimental sample, but we include profile approval as a control to see if it affects the estimates.
In this analysis, we find that the treatment effect on the  number of hires is slightly smaller than in the analysis conditional on platform approval---conditioning the sample on only jobseekers whose profiles were approved has an estimate of \contractsTEpercent{}\% while it is 10\% in the full sample.
The effect on the probability of any hire is \hiredTEpercent \% in the sample of only approved jobseekers and 8\% in the unconditional sample.
This approach and narrowing the sample to only approved jobseekers would ``block'' the approval channel.
In Appendix Table~\ref{tab:main_outcomes_uncon} we report the same analysis not conditioned on profile approval.
None of these robustness checks change the direction or significance of any of the hiring estimates, and the slightly larger estimates in the unconditional sample are unsurprising because of the positive effect of the treatment on platform approval.

\newpage

\section{A simple model of the \clarityView{} of resume writing} \label{sec:discussion}

In this section, we formalize a rational model of how the writing intervention could (a) increase hiring but (b) not lead to worse matches.
We formalize the argument that better writing allowed employers to better ascertain who was a potential match with a simple model, and show how this kind of interplay between resume quality and hiring could exist in equilibrium.

\subsection{A mass of jobseekers with heterogeneous productivity}
There is a unit mass of jobseekers.
If hired, their productivity is $\theta_i$.
Workers are either high-type ($\theta = \theta_H$) or low-type ($\theta = \theta_L$), with $\theta_H > \theta_L$.
Workers know their own type.
It is common knowledge that the fraction of high types in the market is $\gamma$.
All workers, if hired, are paid their expected productivity, from the employer's point of view.
Hires only last one unit of time.

\subsection{Jobseekers decide whether to put into resume-writing}
Before being hired, jobseekers write resumes.
Jobseekers must decide whether to put effort $e \in \{0,1\}$ into writing that resume.
Effort itself is not observable.
The cost of this effort is jobseekers-specific and there is a distribution of individual resume effort costs.
The support of the cost distribution is $[0, \bar{c}]$.
The distribution has mass everywhere and the CDF is $F$ and PDF is $f$.
Jobseekers who put in no effort have resume costs of $0$, while those that put in effort have a cost of $c_i$.
Critically, this cost is independent of a jobseeker's type i.e., there is no Spence-like assumption that better workers find it cheaper to create better resumes \citep{spence1978job}.

Before making an offer, firms observe a signal of jobseekers' type on their resume, $ R \in \{0,1\}$.
With effort, a high-type jobseeker generates an $R = 1$ signal; without effort, $R = 0$.
A low-type jobseeker generates $R = 0$ no matter what.

Clearly, low-types will never put in effort.
The question is whether a high type will put in effort.
The decision hinges on whether the cost of resume effort is worth the wage premium it creates.
Let $w_{R = 0}$ be the wage paid in equilibrium to jobseekers with $R = 0$.
Note that $w_{R = 1} = \theta_H$, as there is no uncertainty about a jobseeker's type if $R = 1$.

A jobseekers $i$ who is a high-type will choose $e = 1$ if $\theta_H - w_{R = 0}(\hat{c}) > c_i$, and so the marginal high-type indifferent between putting in effort or not has a resume-writing cost of $\bar{c}$, where
\begin{align}
  \hat{c} = \theta_H - w_{R = 0}(\hat{c}).
\end{align}
This implies that there are $F(\bar{c}) \gamma$ jobseekers that choose $e = 1$.
There are the high-type jobseekers with relatively low resume writing costs.
The remaining $[1 - F(\bar{c})] \gamma$ high-type jobseekers choose $e = 0$.
They are pooled together with the $1 - \gamma$ jobseekers that choose $e = 0$ because they are low-types.

From the employer's perspective, if they believe that the resume effort cost of the marginal high-type jobseekers is $c$, the probability an $R = 0$ jobseekers is high-type is
\begin{align}
  p_H^{R = 0}(c) = \frac{1 - F(c)}{1/\gamma - F(c)}.
\end{align}
The wage received by an $R = 0$ worker is
\begin{align}
  w_{R = 0}(c) &= \theta_L + (\theta_H - \theta_L) p_H^{R = 0}(c)
\end{align}

When the $c$ of the marginal jobseeker is higher, more jobseekers find it worth choosing $e = 1$, as $F'(c) > 0$.
This leaves fewer high-types in the $R = 0$ pool, and so
\begin{align} \label{eq:dpdc}
\frac{d p_H^{R = 0}}{dc} < 0.
\end{align}

\subsection{The equilibrium fraction of high-type workers putting effort into resume-writing}
In equilibrium, there is some marginal high-type jobseeker indifferent between $e = 0$ and $e = 1$, and so
\begin{align}
  (\theta_H - \theta_L)(1 -  p_H^{R = 0}(\hat{c})) = \hat{c} \nonumber.
\end{align}

\begin{figure}
  \caption{Equilibrium determination of the marginal high-type jobseeker indifferent between putting effort into a resume} \label{images:equil_diagram}
\begin{tikzpicture}
		\begin{axis}[
			width=10 cm,
			axis x line=middle, axis y line=middle,
			axis line style = thick,
			grid style={gray!30},
			ymin=0, ymax=8,
			xmin=0, xmax=8,
			ticks=none,
			]
		\end{axis}
		\begin{scope}
			\node[] at (-0.5, 6)  (n1)    {$\theta_{H}$};
			\node[] at (-1.35, 4.75)  (n2)    {$\theta_{H}\gamma + (1-\gamma)\theta_{L}$};
			\node[] at (8, -1)  (n3)    {$\theta_{H} - c=$};
			\node[] at (9.25, 0)  (n4)    {$\substack{\text{Cost to}\\\text{Marginal}\\\text{Worker}}$};
			\node[] at (8.15, 2)  (n5)    {$\theta_{L}$};
			\node[below=2pt] at (0,0)  (OO)    {$ 0 $};				
			\node[] at (0,0)  (O1)    {};
			\node[] at (9,0)  (O2)    {};%
			\node[] at (-0.25, 6)  (t1)    {};
			\node[] at (0.25, 6)  (t2)    {};
			\node[] at (-0.25, 4.75)  (t3)    {};
			\node[] at (0.25, 4.75)  (t4)    {};
			\node[] at (7.5, 2)  (t5)    {};
			\node[] at (8, 2)  (t6)    {};
			\node[] at (7.75,0.25)  (t7)    {};
			\node[] at (7.75,-0.25)  (t8)    {};
		\end{scope}
		\begin{scope}
			\draw[dashed, black] (7.75,6) -- (7.75,0) node [below=2pt] {$ \bar{c}$};;
			\draw[very thick, name path=SL1] (0,6) -- (7.25, -1);
			\draw[very thick, name path=BL1] (0,4.75) to[bend right=20] (7.75,2);
			\draw[thick, gray, name path=BL2] (0,4.75) to[bend right=27.5] (7.75,2);
			\path [name intersections={of=SL1 and BL1,by=K1}];
			\path [name intersections={of=SL1 and BL2,by=K2}];
			\draw [dashed] (K1) -- ($(O1)!(K1)!(O2)$) node [below=2pt] {$ \hat{c} $};
			\draw [dashed, gray] (K2) -- ($(O1)!(K2)!(O2)$) node [below=2pt] {\color{gray}$ \hat{c}' $};%
			\draw[->,thick, red] (5.85,4.5) to[bend right=20] (4.5,2.05);
			\draw[->,thick, red] (8.35,2.75) to[bend right=20] (7.75,2);
			\draw[very thick] (t1) to (t2);
			\draw[very thick] (t3) to (t4);
			\draw[very thick] (t5) to (t6);
			\draw[very thick] (t7) to (t8);
		\end{scope}%
		\begin{scope}
			\node[mybox] at (6, 5) {\scriptsize Resume writing costs decrease};
			\node[mybox] at (9.5,3.5) {\scriptsize Payoff to marginal H-type worker when $ R=0 $};
			\node[mybox] at (8,-1.75) {\scriptsize Payoff to marginal H-type worker when $ R=1 $};
		\end{scope}%
	\end{tikzpicture}
   \end{figure}
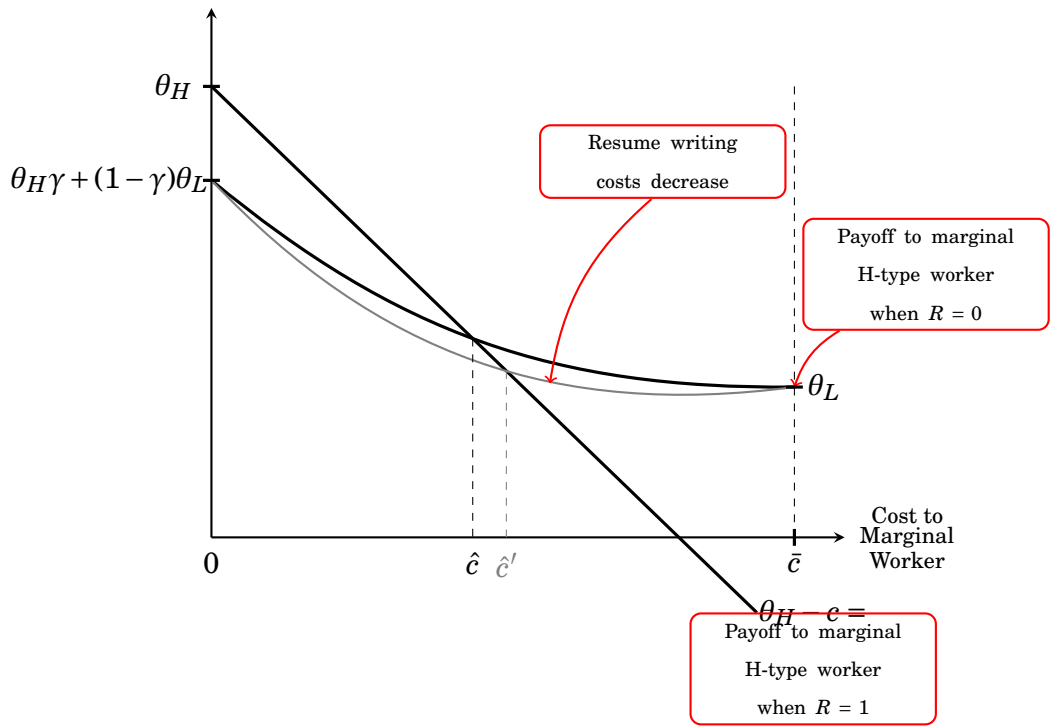

Figure~\ref{images:equil_diagram} illustrates the equilibrium i.e., the cost where the marginal jobseeker is indifferent between $e = 0$ and $e = 1$.
The two downward-sloping lines are the pay-offs to the marginal jobseeker for each $c$.
The pay-off to $R = 1$ is declining, as the wage is constant (at $\theta_H$) but the cost is growing linearly.
The pay-off to $R = 0$ is also declining, from Equation~\ref{eq:dpdc}.
Both curves are continuous.

Note that when the marginal jobseeker has $c = 0$, there is just a point-mass of high-types that have a cost that low, i.e., $f(c)$.
This implies that the $R = 0$ pool is just the expected value of all jobseekers, so the wage is just $\gamma \theta_H + (1-\gamma)\theta_L$.
The ``marginal'' jobseeker pays nothing, so the pay-off is $\theta_H$.
At the other extreme, $c = \bar{c}$, all but a point mass of jobseekers have a cost less than this, so the $R = 0$ pool is purely low-types and the wage is $\theta_L$.
For the $R = 1$ market, the ``marginal'' jobseeker has a cost of $\bar{c}$ so the pay-off is $\theta_H - \bar{c}$.
We know $\theta_H > \gamma \theta_H + (1-\gamma) \theta_L$.
And by assumption, $\theta_L > \theta_H - \bar{c}$, and so by the intermediate value theorem, an equilibrium $\hat{c}$ exists on $(0,\bar{c})$.

\subsection{A shift in the resume writing cost distribution leads to more high-type workers choosing to exert effort}
Now suppose a technology comes along that lowers---or at least keeps the same---resume writing costs for all jobseekers.
This would shift $F$ higher for all points except the endpoints of the support, creating a new distribution of costs that first-order stochastically dominates the other.

Before determining the new equilibrium, note that no matter the marginal $c$, when $F$ increases, the probability that an $R = 0$ worker is a high-type declines, as
\begin{align}
  \frac{d p_H}{d F} = -\frac{1}{(F - 2)^2} < 0.
\end{align}
This shifts the $w_{R = 0}$ curve down everywhere, without changing the endpoints.

Because $w_{R = 1} - c$ is downward sloping, it intersects $w_{R = 0}(c)$ at a higher value of $c$.
At the new equilibrium, the marginal jobseeker has resumes costs of $\hat{c}'$, where $\hat{c}' > \hat{c}$.
At this new equilibrium, more jobseekers choose $e = 1$, causing more $R = 1$ signals.
This lowers wages for the $R = 0$ group.

\subsection{The effects of lower costs are theoretically ambiguous}
Note that this shift in costs is not Pareto improving---low-types are made worse off as they find themselves in a pool with fewer high-types.
Furthermore, because workers are all paid their expected product, the ideal outcome would be for everyone to choose $R = 0$.
Resume effort purely changes around the allocation of the wage bill, not the total amount.
Total surplus is
\begin{align}
  \theta_H \gamma + (1-\gamma) \theta_L - \int_0^{\bar{c}} c f(c) dc,
\end{align}
which is maximized at $\hat{c} = 0$, i.e., when no one finds it worthwhile to choose effort.
However, with a \emph{shift} in cost distribution (raising $F$), whether matters is whether the marginal decrease in costs for all inframarginal workers i..e, those with $c < \hat{c}$ outweighs the costs borne by the (newly) marginal jobseekers who choose to put in effort.

In our model, all job offers are accepted.
However, if we think of jobseekers as having idiosyncratic reservation values that determine whether they accept an offer, the shift in costs makes it more likely that high-types will accept an offer, while making it less likely that low-types will accept an offer.
This is consistent with our results of a greater chance an employer hires at all in the treatment.
It is also consistent with our result of higher wages.
Finally, if we think of employer ratings being a function of surplus, our finding of no change in satisfaction is also consistent, as employers are, in all cases, just paying for expected productivity.

\section{Conclusion}\label{sec:conclusion}

Employers are more likely to hire new labor market entrants with better-written resumes.
We argue that better writing makes it easier for employers to decide to hire a particular worker.
We show results from a field experiment in an online labor market where treated workers were given algorithmic writing assistance from \Grammarly{}.
These jobseekers were \hiredTEpercent \% more likely to get hired and formed \contractsTEpercent \% more contracts over the month-long experiment.
While one might have expected writing quality to be a valuable indicator of worker quality, the treatment did not affect employers' ratings of hired workers.
We provide a model of the hiring process where the cost of exerting effort on a resume is lowered by algorithmic writing assistance, which helps employers to distinguish between high and low-type workers.

 One possibility is that the benefits to treated workers came at the expense of other workers, as both treated- and control-assigned workers compete in the same market.
 Crowd-out concerns have been shown to be important with labor market assistance~\citep{crepon2013labor}.
 However, even if additional hires came from experienced workers, this is likely still a positive result.
 New labor market entrants are uniquely disadvantaged \citep{pallais2010inefficient} in online labor markets.
 To the extent that the gains to new workers come partially at the expense of experienced workers, this is likely a good trade-off.

Conceptualizing AI/ML innovation and proliferation as a fall in the cost of prediction technology fits our setting \citep{agrawal2018predictionmachines, agrawal2018prediction}.
Writing a resume is, in part, an applied prediction task---what combination of words and phrases, arranged in what order, are likely to maximize my pay-off from a job search?
\Grammarly{} reduces the effort or cost required for making these decisions. When revising their resumes, rather than identifying errors in their own predictions themselves, jobseekers with access to \Grammarly{}  specify their target audience and writing goals and enter their draft profiles into \Grammarly.  \Grammarly{} assists jobseekers in error correction.
Furthermore, the treatment, by lowering the costs of error-free writing for at least some jobseekers, causes them to do better at writing their resumes.

Interestingly, this algorithmic writing assistance will likely ``ruin'' writing as a signal.
With the proliferation of writing technologies with capabilities far beyond what is explored here~\citep{gpt3}, even if the \signalView{} was at one time true, technological changes are likely to make it not true in the future.

\bibliographystyle{aer}
\bibliography{grammarly.bib}

\newpage
\appendix
\section{Appendix}

\subsection{Additional Tables and Figures} \label{sec:appendix}

\input{tables/approved.tex}

\begin{figure}
	\caption{Daily allocations of jobseekers into experimental cells}
         \includegraphics[width=\textwidth]{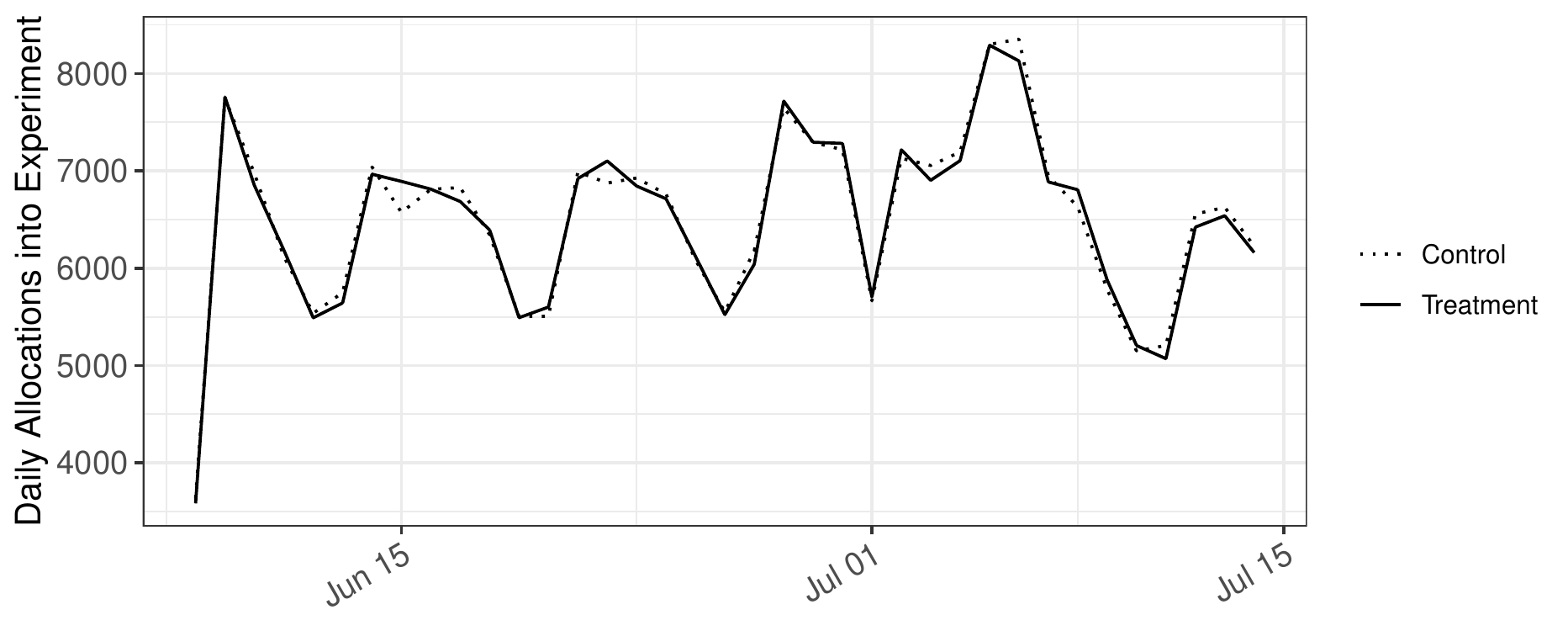}
         \label{fig:allocations}
             \emph{ \footnotesize Notes: This plot shows the daily allocations into the treatment and control cells for the experimental sample of~\numberTotal{} new jobseekers to the platform.}
\end{figure}

\input{tables/length.tex}

\newpage

\input{tables/main_outcomes.tex}

\input{tables/main_outcomes_uncon.tex}

\input{tables/main_outcomes_control.tex}

\input{images/languagetool_description.tex}

\newpage

\begin{table}[h!]
\caption{Summary statistics on error counts and rates in the control group \label{tab:error_summary_stats}}

\begin{center}
\begin{tabular}{llll}
\toprule\\
& &   Total Errors &     Error Rate
\\
\hline
  \input{./tables/summary_languagetool_error.tex} \\ 
\bottomrule
 \end{tabular}
 \end{center}

\begin{footnotesize}
  \emph{Notes}: This table reports means and standard errors of the writing errors in the resumes of the control group. The first column displays the total error count and the second column displays the error rate (normalized by word count). Writing errors are defined by LanguageToolR. The sample is made up of all jobseekers in the control group of the experimental sample who submitted non-empty resumes and were approved by the platform.
\end{footnotesize}
\end{table}

\input{tables/predict_hiring_languagetool}
\begin{landscape}
\input{tables/predict_hiring_languagetool_total}
\end{landscape}
\begin{landscape}
\input{tables/writing_outcomes_1.tex}
\end{landscape}
\begin{landscape}
\input{tables/writing_outcomes_2.tex}
\end{landscape}

\end{document}

%% file: parameter_files/parameters.tex
\newcommand{\numberTreat}{240,231}
\newcommand{\numberControl}{240,717}
\newcommand{\percentTreat}{ 50}
\newcommand{\numberTotal}{480,948}

\newcommand{\numberApprovedNonEmpty}{194,700}
\newcommand{\meanLengthNoErrors}{52}

\newcommand{\meanCorrectRate}{96}
\newcommand{\percentOver}{ 98}
\newcommand{\hiredTEpercent}{8}
\newcommand{\hiredControl}{3}
\newcommand{\contractsTEpercent}{7.8}

\newcommand{\bidsControl}{24}

\newcommand{\wagesTEpercent}{8.4}
\newcommand{\wagesControl}{17.17}
\newcommand{\wagesTreatment}{18.62}
\newcommand{\wagesPvalue}{0.059}
\newcommand{\privateRatingControl}{8.835}
\newcommand{\privateRatingTreat}{8.84}

\newcommand{\percentProfilesSubmitted}{46}
\newcommand{\numProfilesSubmitted}{219,242}
\newcommand{\numProfilesSubmittedControl}{109,604}
\newcommand{\numProfilesSubmittedTreat}{109,638}

\newcommand{\percentAllProfilesApproved}{41}
\newcommand{\numSubProfilesApproved}{195,469}
\newcommand{\numSubProfilesApprovedControl}{97,610}
\newcommand{\numSubProfilesApprovedTreat}{97,859}

\newcommand{\numProfilesNonEmpty}{194,700}
\newcommand{\numProfilesNonEmptyControl}{97,221}
\newcommand{\numProfilesNonEmptyTreat}{97,479}

%% file: tables/summary_stats_post.tex
\\
  \multicolumn{3}{l}{\emph{Full sample description: N = 480,948}}\\
& Resume submitted & 0.456 (0.001) & 0.455 (0.001) & 0.001 (0.001) & 0.454 &   \\ 
 
& Platform approved & 0.407 (0.001) & 0.406 (0.001) & 0.002 (0.001) & 0.187 &   \\ 
 
& Resume length & 32.910 (0.116) & 32.859 (0.117) & 0.051 (0.165) & 0.758 &   \\ 
 
& Profile hourly rate & 18.843 (0.126) & 18.917 (0.126) & -0.075 (0.178) & 0.676 &   \\ 
 
\\

%% file: tables/summary_stats_pre.tex
\\
 \multicolumn{3}{l}{\emph{Pre-allocation attributes of the analysis sample: N = 194,700}}\\
& From English-speaking country & 0.182 (0.001) & 0.183 (0.001) & -0.002 (0.002) & 0.363 &   \\ 
 
& US-based & 0.141 (0.001) & 0.143 (0.001) & -0.002 (0.002) & 0.223 &   \\ 
 
& Specializing in writing & 0.166 (0.001) & 0.168 (0.001) & -0.002 (0.002) & 0.150 &   \\ 
 
& Specializing in software & 0.115 (0.001) & 0.115 (0.001) & 0.000 (0.001) & 0.769 &   \\ 
 
& Resume length & 70.393 (0.222) & 70.260 (0.222) & 0.133 (0.314) & 0.671 &   \\ 
 
\\

%% file: tables/predict_hiring_languagetool_perword.tex
\begin{table}[!htbp] \centering 
  \caption{Hiring outcomes predicted based on language errors (normalized by word count) in the control group } 
  \label{tab:predict_hiring_languagetool_perword} 
\small 
\begin{tabular}{@{\extracolsep{5pt}}lcc} 
\\[-1.8ex]\hline 
\hline \\[-1.8ex] 
 & \multicolumn{2}{c}{\textit{Dependent variable:}} \\ 
\cline{2-3} 
\\[-1.8ex] & Number of Contracts & Hired \\ 
\\[-1.8ex] & (1) & (2)\\ 
\hline \\[-1.8ex] 
 Capitalization Error & $-$0.298$^{***}$ & $-$0.175$^{***}$ \\ 
  & (0.077) & (0.039) \\ 
  Possible Typo & $-$0.049$^{***}$ & $-$0.031$^{***}$ \\ 
  & (0.011) & (0.006) \\ 
  Grammar Error & $-$0.320$^{***}$ & $-$0.223$^{***}$ \\ 
  & (0.097) & (0.049) \\ 
  Punctuation Error & 0.064$^{***}$ & 0.038$^{***}$ \\ 
  & (0.024) & (0.012) \\ 
  Typography Error & $-$0.053$^{***}$ & $-$0.039$^{***}$ \\ 
  & (0.016) & (0.008) \\ 
  Style Error & 0.164$^{*}$ & 0.092$^{*}$ \\ 
  & (0.098) & (0.050) \\ 
  Miscellaneous Error & $-$0.457$^{***}$ & $-$0.241$^{***}$ \\ 
  & (0.143) & (0.073) \\ 
  Redundant Phrases & 0.123 & 0.086 \\ 
  & (0.385) & (0.195) \\ 
  Nonstandard Phrases & 0.860 & $-$0.074 \\ 
  & (1.275) & (0.646) \\ 
  Commonly Confused Words & $-$1.192$^{**}$ & $-$0.667$^{**}$ \\ 
  & (0.531) & (0.269) \\ 
  Collocations & $-$0.588$^{*}$ & $-$0.368$^{**}$ \\ 
  & (0.347) & (0.176) \\ 
  Semantic Error & $-$1.229 & $-$0.683$^{*}$ \\ 
  & (0.789) & (0.400) \\ 
  Constant & 0.167 & 0.167$^{**}$ \\ 
  & (0.142) & (0.072) \\ 
 \hline \\[-1.8ex] 
Controls & X & X \\ 
Observations & 93,725 & 93,725 \\ 
R$^{2}$ & 0.002 & 0.004 \\ 
\hline 
\hline \\[-1.8ex] 
\end{tabular}
\\
\begin{minipage}{\textwidth}
{\footnotesize \emph{Notes}: This table analyzes correlation between various writing errors on jobseekers' resumes and their hiring outcomes. 
          The independent variables, writing errors, are divded by the number of words in the jobseekers' resume.
          Column (1) defines Number of Contracts as the number of unique jobs they work over the month after they register for the platform.
          Column (2) defines Hired as the probability the jobseeker was hired over that month.
          All analysis includes controls for profile hourly rate and job category. Writing errors are defined by LanguageToolR. 
           The sample is made up of all jobseekers in the control group of the experimental sample who submitted non-empty resumes and were approved by the platform. 
          \newline \starlanguage}
\end{minipage}
\end{table}

%% file: tables/treatment.tex
\begin{table}[!htbp] \centering 
  \caption{Effects of writing assistance on spelling} 
  \label{tab:treatment} 
\small 
\begin{tabular}{@{\extracolsep{5pt}}lcccc} 
\\[-1.8ex]\hline 
\hline \\[-1.8ex] 
 & \multicolumn{4}{c}{\textit{Dependent variable:}} \\ 
\cline{2-5} 
\\[-1.8ex] & \multicolumn{4}{c}{Frac Words Spelled Correctly x 100} \\ 
\\[-1.8ex] & (1) & (2) & (3) & (4)\\ 
\hline \\[-1.8ex] 
 Algo Writing Treatment (Trt) & 0.090$^{*}$ & 0.133$^{***}$ & 0.122$^{**}$ & 0.102$^{**}$ \\ 
  & (0.046) & (0.051) & (0.050) & (0.050) \\ 
  Native-English &  & 2.405$^{***}$ &  &  \\ 
  &  & (0.084) &  &  \\ 
  Trt $\times $English &  & $-$0.218$^{*}$ &  &  \\ 
  &  & (0.119) &  &  \\ 
  US &  &  & 2.394$^{***}$ &  \\ 
  &  &  & (0.093) &  \\ 
  Trt $\times$ US &  &  & $-$0.192 &  \\ 
  &  &  & (0.131) &  \\ 
  Writer &  &  &  & 0.875$^{***}$ \\ 
  &  &  &  & (0.087) \\ 
  Trt $\times$ Writer &  &  &  & $-$0.063 \\ 
  &  &  &  & (0.123) \\ 
  Constant & 96.399$^{***}$ & 95.957$^{***}$ & 96.055$^{***}$ & 96.251$^{***}$ \\ 
  & (0.033) & (0.036) & (0.035) & (0.036) \\ 
 \hline \\[-1.8ex] 
Observations & 194,700 & 194,700 & 194,700 & 194,700 \\ 
R$^{2}$ & 0.00002 & 0.008 & 0.006 & 0.001 \\ 
\hline 
\hline \\[-1.8ex] 
\end{tabular}
\\
\begin{minipage}{ \textwidth}
{\footnotesize \emph{Notes}:
           In Column (1) we show the overall effect of the treatment to the percentage of correctly spelled words on a jobseekers' resume.
          In Column (2) we interact the treatment with a dummy variable for if the jobseeker is in the US, UK, Canada, or Australia.
          In Column (3) we interact the treatment with a dummy for if the jobseeker is in the US.
          In Column (4) we interact the treatment with a dummy for if the jobseeker lists Writing as their primary category of desired work.
          The sample is made up of all jobseekers in the control group of the experimental sample who submitted non-empty resumes and were approved by the platform. 
          \starlanguage}
\end{minipage}
\end{table}

%% file: tables/applications.tex
\begin{table}[!htbp] \centering 
  \caption{Effects of writing assistance on jobseekers' application behavior} 
  \label{tab:applications} 
\small 
\begin{tabular}{@{\extracolsep{5pt}}lccc} 
\\[-1.8ex]\hline 
\hline \\[-1.8ex] 
 & \multicolumn{3}{c}{\textit{Dependent variable:}} \\ 
\cline{2-4} 
\\[-1.8ex] & Num Applications & Num Hourly Applications & Mean Hourly Wage Bid \\ 
\\[-1.8ex] & (1) & (2) & (3)\\ 
\hline \\[-1.8ex] 
 Algo Writing Treatment & 0.010 & 0.009 & $-$0.223 \\ 
  & (0.028) & (0.018) & (0.391) \\ 
  Constant & 2.466$^{***}$ & 1.325$^{***}$ & 24.283$^{***}$ \\ 
  & (0.020) & (0.012) & (0.276) \\ 
 \hline \\[-1.8ex] 
Observations & 194,700 & 194,700 & 68,664 \\ 
R$^{2}$ & 0.00000 & 0.00000 & 0.00000 \\ 
\hline 
\hline \\[-1.8ex] 
\end{tabular}
\\
\begin{minipage}{ \textwidth}
{\footnotesize \emph{Notes}: 
          This table analyzes the effect of the treatment on jobseekers' application behavior.
          The outcome in Column (1) is the number of total applications a jobseeker sent out between the time the experiment began and one month after it ended.
           The sample is made up of all jobseekers in the control group of the experimental sample who submitted non-empty resumes and were approved by the platform.
          The outcome in Column (2) is the number of specifically hourly applications sent out in that same time period.
          The outcome in Column (3) is the mean hourly wage bid they proposed for those hourly jobs, and the sample narrows to 
          only jobseeker who submitted at least one application to an hourly job.
          
          \starlanguage}
\end{minipage}
\end{table}

%% file: tables/hiring.tex
\begin{table}[!htbp] \centering 
  \caption{Effects of writing assistance on hiring, by sub-groups} 
  \label{tab:hiring} 
\small 
\begin{tabular}{@{\extracolsep{5pt}}lcccc} 
\\[-1.8ex]\hline 
\hline \\[-1.8ex] 
 & \multicolumn{4}{c}{\textit{Dependent variable:}} \\ 
\cline{2-5} 
\\[-1.8ex] & \multicolumn{4}{c}{Hired x 100} \\ 
\\[-1.8ex] & (1) & (2) & (3) & (4)\\ 
\hline \\[-1.8ex] 
 Algo Writing Treatment (Trt) & 0.247$^{***}$ & 0.223$^{**}$ & 0.242$^{***}$ & 0.237$^{***}$ \\ 
  & (0.080) & (0.088) & (0.086) & (0.088) \\ 
  Native-English &  & 2.508$^{***}$ &  &  \\ 
  &  & (0.146) &  &  \\ 
  Trt $\times $English &  & 0.155 &  &  \\ 
  &  & (0.207) &  &  \\ 
  US &  &  & 2.602$^{***}$ &  \\ 
  &  &  & (0.161) &  \\ 
  Trt $\times$ US &  &  & 0.072 &  \\ 
  &  &  & (0.228) &  \\ 
  Writer &  &  &  & $-$0.293$^{*}$ \\ 
  &  &  &  & (0.151) \\ 
  Trt $\times$ Writer &  &  &  & 0.061 \\ 
  &  &  &  & (0.214) \\ 
  Constant & 3.093$^{***}$ & 2.632$^{***}$ & 2.719$^{***}$ & 3.142$^{***}$ \\ 
  & (0.057) & (0.063) & (0.061) & (0.062) \\ 
 \hline \\[-1.8ex] 
Observations & 194,700 & 194,700 & 194,700 & 194,700 \\ 
R$^{2}$ & 0.00005 & 0.003 & 0.003 & 0.0001 \\ 
\hline 
\hline \\[-1.8ex] 
\end{tabular}
\\
\begin{minipage}{ \textwidth}
{\footnotesize \emph{Notes}:  This table analyzes the effect of the treatment on whether or not a jobseeker was ever hired on the platform in the month after they joined, times 100.
          In Column (1) we show the overall effect of the treatment to hiring.
          In Column (2) we interact the treatment with a dummy variable for if the jobseeker is in the US, UK, Canada, or Australia.
          In Column (3) we interact the treatment with a dummy for if the jobseeker is in the US.
          In Column (4) we interact the treatment with a dummy for if the jobseeker lists Writing as their primary category of desired work.
          The sample is made up of all jobseekers in the control group of the experimental sample who submitted non-empty resumes and were approved by the platform. 
          \starlanguage}
\end{minipage}
\end{table}

%% file: tables/wages_and_rates.tex
\begin{table}[!htbp] \centering 
  \caption{Effect of algorithmic writing assistance on wages and ratings of worked jobs} 
  \label{tab:wages_and_rates} 
\small 
\begin{tabular}{@{\extracolsep{5pt}}lcc} 
\\[-1.8ex]\hline 
\hline \\[-1.8ex] 
 & \multicolumn{2}{c}{\textit{Dependent variable:}} \\ 
\cline{2-3} 
\\[-1.8ex] & Hourly wage rate & Private rating \\ 
\\[-1.8ex] & (1) & (2)\\ 
\hline \\[-1.8ex] 
 Algo Writing Treatment & 1.448$^{*}$ & 0.004 \\ 
  & (0.766) & (0.072) \\ 
  Constant & 17.173$^{***}$ & 8.835$^{***}$ \\ 
  & (0.558) & (0.052) \\ 
 \hline \\[-1.8ex] 
Observations & 3,542 & 4,433 \\ 
R$^{2}$ & 0.001 & 0.00000 \\ 
\hline 
\hline \\[-1.8ex] 
\end{tabular}
\\
\begin{minipage}{0.7 \textwidth}
{\footnotesize \emph{Notes}: 
        This analysis looks at the effect of treatment on outcomes of worked jobs for jobseekers in the experimental sample.
         Column (1) defines hourly wage rate as the hourly wage rate paid for all hourly jobs worked.
          Column (2) defines private rating as the private rating on all jobs given by employers to the workers after the job ended.
        The experimental sample is of all new jobseekers who registered and were approved for the platform between June 8th and July 14th, 2021 
        and had non-empty resumes.
          \starlanguage}
\end{minipage}
\end{table}

%% file: tables/approved.tex
\begin{table}[!htbp] \centering 
  \caption{Effects of writing assistance on profile submission and platform approval} 
  \label{tab:approved} 
\small 
\begin{tabular}{@{\extracolsep{5pt}}lccc} 
\\[-1.8ex]\hline 
\hline \\[-1.8ex] 
 & \multicolumn{3}{c}{\textit{Dependent variable:}} \\ 
\cline{2-4} 
\\[-1.8ex] & Profile submitted x 100 & \multicolumn{2}{c}{Approved x 100} \\ 
\\[-1.8ex] & (1) & (2) & (3)\\ 
\hline \\[-1.8ex] 
 Algo Writing Treatment & 0.106 & 0.199 & 0.186 \\ 
  & (0.144) & (0.133) & (0.142) \\ 
  Constant & 45.532$^{***}$ & 89.057$^{***}$ & 40.550$^{***}$ \\ 
  & (0.102) & (0.094) & (0.100) \\ 
 \hline \\[-1.8ex] 
Observations & 480,948 & 219,242 & 480,948 \\ 
R$^{2}$ & 0.00000 & 0.00001 & 0.00000 \\ 
\hline 
\hline \\[-1.8ex] 
\end{tabular}
\\
\begin{minipage}{ \textwidth}
{\footnotesize \emph{Notes}: 
          This table analyzes the effect of the treatment on 
          In Column (1) the outcome is 100 times a binary indicator for whether or not the jobseeker completed platform registration and submitted their resume, on the full experimental sample.
          In Column (2) the outcome is 100 times a binary indicator for whether or not the platform approved the resume, on the sample of only those jobseekers who submitted their resumes. 
          In Column (3) the outcome is 100 times a binary indicator for whether or not the platform approved the resume, on the full experimental sample.
          \starlanguage}
\end{minipage}
\end{table}

%% file: tables/length.tex
\begin{table}[!htbp] \centering 
  \caption{Effects of writing assistance on length of resume} 
  \label{tab:length} 
\small 
\begin{tabular}{@{\extracolsep{5pt}}lc} 
\\[-1.8ex]\hline 
\hline \\[-1.8ex] 
 & \multicolumn{1}{c}{\textit{Dependent variable:}} \\ 
\cline{2-2} 
\\[-1.8ex] & Number of words in resume \\ 
\hline \\[-1.8ex] 
 Algo Writing Treatment & 0.127 \\ 
  & (0.314) \\ 
  Constant & 70.541$^{***}$ \\ 
  & (0.223) \\ 
 \hline \\[-1.8ex] 
Observations & 194,700 \\ 
R$^{2}$ & 0.00000 \\ 
\hline 
\hline \\[-1.8ex] 
\end{tabular}
\\
\begin{minipage}{0.7 \textwidth}
{\footnotesize \emph{Notes}: 
           This table analyzes the effect of the treatment on the number of words in a jobseeker's resume. 
         The sample is made up of all jobseekers in the experimental sample who submitted non-empty profiles and were approved by the platform. \starlanguage}
\end{minipage}
\end{table}

%% file: tables/main_outcomes.tex
\begin{table}[!htbp] \centering 
  \caption{Effect of algorithmic writing assistance on hiring outcomes} 
  \label{tab:main_outcomes} 
\small 
\begin{tabular}{@{\extracolsep{5pt}}lcccc} 
\\[-1.8ex]\hline 
\hline \\[-1.8ex] 
 & \multicolumn{4}{c}{\textit{Dependent variable:}} \\ 
\cline{2-5} 
\\[-1.8ex] & Num Contracts & Hired x 100 & Num Hourly Interviews & Num Invitations \\ 
\\[-1.8ex] & (1) & (2) & (3) & (4)\\ 
\hline \\[-1.8ex] 
 Algo Writing Treatment & 0.004$^{**}$ & 0.247$^{***}$ & 0.002 & 0.001 \\ 
  & (0.002) & (0.080) & (0.004) & (0.003) \\ 
  Constant & 0.047$^{***}$ & 3.093$^{***}$ & 0.210$^{***}$ & 0.142$^{***}$ \\ 
  & (0.001) & (0.057) & (0.003) & (0.002) \\ 
 \hline \\[-1.8ex] 
Observations & 194,700 & 194,700 & 194,700 & 194,700 \\ 
R$^{2}$ & 0.00003 & 0.00005 & 0.00000 & 0.00000 \\ 
\hline 
\hline \\[-1.8ex] 
\end{tabular}
\\
\begin{minipage}{ \textwidth}
{\footnotesize \emph{Notes}: 
        This analysis looks at the effect of treatment on hiring outcomes on jobseekers in the experimental sample.
         Column (1) defines Number of Contracts as the number of unique jobs they work over the month after they register for the platform.
          Column (2) defines Hired x 100 as  one hundred times the probability the jobseeker was hired over that month. 
          Column (3) is the number of interviews they gave over that month. 
          And the Column (4) outcome Invitations is the number of times they were recruited to a job over their first month.
        The experimental sample is of all new jobseekers who registered and were approved for the platform between June 8th and July 14th, 2021 
        and had non-empty resumes.
          \starlanguage}
\end{minipage}
\end{table}

%% file: tables/main_outcomes_uncon.tex
\begin{table}[!htbp] \centering 
  \caption{Effect of algorithmic writing assistance on hiring outcomes, unconditional on platform approval} 
  \label{tab:main_outcomes_uncon} 
\small 
\begin{tabular}{@{\extracolsep{5pt}}lcccc} 
\\[-1.8ex]\hline 
\hline \\[-1.8ex] 
 & \multicolumn{4}{c}{\textit{Dependent variable:}} \\ 
\cline{2-5} 
\\[-1.8ex] & Num Contracts & Hired x 100 & Num Hourly Interviews & Num Invitations \\ 
\\[-1.8ex] & (1) & (2) & (3) & (4)\\ 
\hline \\[-1.8ex] 
 Algo Writing Treatment & 0.002$^{**}$ & 0.110$^{***}$ & 0.001 & 0.001 \\ 
  & (0.001) & (0.033) & (0.002) & (0.001) \\ 
  Constant & 0.019$^{***}$ & 1.273$^{***}$ & 0.085$^{***}$ & 0.058$^{***}$ \\ 
  & (0.0005) & (0.023) & (0.001) & (0.001) \\ 
 \hline \\[-1.8ex] 
Observations & 480,948 & 480,948 & 480,948 & 480,948 \\ 
R$^{2}$ & 0.00001 & 0.00002 & 0.00000 & 0.00000 \\ 
\hline 
\hline \\[-1.8ex] 
\end{tabular}
\\
\begin{minipage}{ \textwidth}
{\footnotesize \emph{Notes}: 
        This analysis looks at the effect of treatment on hiring outcomes on jobseekers in the experimental sample.
         Column (1) defines Number of Contracts as the number of unique jobs they work over the month after they register for the platform.
          Column (2) defines Hired x 100 as  one hundred times the probability the jobseeker was hired over that month. 
          Column (3) is the number of interviews they gave over that month. 
          And the Column (4) outcome Invitations is the number of times they were recruited to a job over their first month.
        The sample used in this analysis is the entire experimental sample.
          \starlanguage}
\end{minipage}
\end{table}

%% file: tables/main_outcomes_control.tex
\begin{table}[!htbp] \centering 
  \caption{Effect of algorithmic writing assistance on hiring outcomes, controlling for platform approval} 
  \label{tab:main_outcomes_control} 
\small 
\begin{tabular}{@{\extracolsep{5pt}}lcccc} 
\\[-1.8ex]\hline 
\hline \\[-1.8ex] 
 & \multicolumn{4}{c}{\textit{Dependent variable:}} \\ 
\cline{2-5} 
\\[-1.8ex] & Num Contracts & Hired x 100 & Num Hourly Interviews & Num Invitations \\ 
\\[-1.8ex] & (1) & (2) & (3) & (4)\\ 
\hline \\[-1.8ex] 
 Algo Writing Treatment & 0.002$^{**}$ & 0.104$^{***}$ & 0.001 & 0.0004 \\ 
  & (0.001) & (0.033) & (0.002) & (0.001) \\ 
  Approved by Platform & 0.048$^{***}$ & 3.171$^{***}$ & 0.210$^{***}$ & 0.142$^{***}$ \\ 
  & (0.001) & (0.033) & (0.002) & (0.001) \\ 
  Constant & $-$0.0003 & $-$0.013 & $-$0.0003 & 0.00001 \\ 
  & (0.001) & (0.027) & (0.001) & (0.001) \\ 
 \hline \\[-1.8ex] 
Observations & 480,948 & 480,948 & 480,948 & 480,948 \\ 
R$^{2}$ & 0.011 & 0.019 & 0.030 & 0.023 \\ 
\hline 
\hline \\[-1.8ex] 
\end{tabular}
\\
\begin{minipage}{ \textwidth}
{\footnotesize \emph{Notes}: 
        This analysis looks at the effect of treatment on hiring outcomes on jobseekers in the experimental sample.
         Column (1) defines Number of Contracts as the number of unique jobs they work over the month after they register for the platform.
          Column (2) defines Hired x 100 as  one hundred times the probability the jobseeker was hired over that month. 
          Column (3) is the number of interviews they gave over that month. 
          And the Column (4) outcome Invitations is the number of times they were recruited to a job over their first month.
       The sample used in this analysis is the entire experimental sample.
          \starlanguage}
\end{minipage}
\end{table}

%% file: images/languagetool_description.tex
\begin{table}[hb]
 \footnotesize
 \caption{Description of Error Rule Categories with Examples}
 \label{tab:languagetooldescription}
\begin{tabular}{p{1in}|p{2.5in}|p{3in}}
\hline
\textbf{Category}                & \textbf{Description}                                                                                                                                                                                                       & \textbf{Examples}                                                                                                                                                                                                                                     \\ \hline
American English Phrases         & Sentence favors the American English spelling of words.                                                                                                                                                                    & \textit{apologize, catalog, civilization, defense}                                                                                                                                                                                                         \\ \hline
British English, Oxford Spelling & Sentence favors the British English spelling of words.                                                                                                                                                                     & \textit{apologise, catalogue, civilisation, defence }                                                                                                                                                                                                           \\ \hline
Capitalization                   & Rules about detecting uppercase words where lowercase is required and vice versa.                                                                                                                                          & \textit{This house is old. it was built in 1950.\newline I really like Harry potter.}                                                                                                                                         \\ \hline
Collocations                     & A collocation is made up of two or more words that are commonly used together in English. This refers to an error in this type of phrase.                                                                                  & \textit{Undoubtedly, this is the result of an extremely dynamic development of Lublin in the recent years.\newline I will take it in to account.\newline It's batter to be save then sorry.}                                \\ \hline
Commonly Confused Words          & Words that are easily confused, like 'there' and 'their' in English.                                                                                                                                                       & \textit{I have my won bed.\newline Their elicit behavior got the students kicked out of school.It's the worse possible outcome.}                                                                                          \\ \hline
Grammar                          & Violations related to system of rules that allow us to structure sentences.                                                                                                                                                & \textit{Tom make his life worse.\newline A study like this one rely on historical and present data.This is best way of dealing with errors.   }                                                                              \\ \hline
Miscellaneous                    & Miscellaneous rules that don't fit elsewhere.                                                                                                                                                                              & \textit{This is best way of dealing with errors.\newline The train arrived a hour ago.\newline It's nice, but it doesn’t work. (inconsistent apostrophes)  }                                                                      \\ \hline
Nonstandard Phrases              &                                                                                                                                                                                                                            & \textit{I never have been to London.\newline List the names in an alphabetical order.\newline Why would a man all of the sudden send flowers? }                                                                                   \\ \hline
Possible Typo                    & Spelling issues.                                                                                                                                                                                                           & \textit{It'a extremely helpful when it comes to homework.\newline We haven't earned anything.This is not a HIPPA violation. }                                                                                              \\ \hline
Punctuation                      & Error in the marks, such as period, comma, and parentheses, used in writing to separate sentences and their elements and to clarify meaning.                                                                               & \textit{"I'm over here, she said.\newline Huh I thought it was done already. \newline The U.S.A is one of the largest countries. }                                                                                                   \\ \hline
Redundant Phrases                & Redundant phrases contain words that say the same thing twice. When one of the words is removed, the sentence still makes sense. Sometimes the sentence has to be slightly restructured, but the message remains the same. & \textit{We have more than 100+ customers.\newline He did it in a terrible way.\newline The money is sufficient enough to buy the sweater.  }                                                                                        \\ \hline
Semantics                        & Logic, content, and consistency problems.                                                                                                                                                                                  & \textit{It allows us to both grow, focus, and flourish.\newline On October 7, 2025 , we visited the client.This was my 11nd try. }                                                                                           \\ \hline
Style                            & General style issues not covered by other categories, like overly verbose wording.                                                                                                                                         & \textit{Moreover, the street is almost entirely residential. Moreover, it was named after a poet.\newline Doing it this way is more easy than the previous method.\newline I'm not very experienced too. \newline Anyways, I don't like it. } \\ \hline
Typography                       & Problems like incorrectly used dash or quote characters.                                                                                                                                                                   & \textit{This is a  sentence with two consecutive spaces.\newline I have 3dogs.The price rose by \$12,50.\newline I'll buy a new T—shirt.   }                                                                                       \\ \hline
\end{tabular}
\end{table}

%% file: tables/summary_languagetool_error.tex
\\
& Capitalization Errors & 0.112 (0.488) & 0.003 (0.015) \\ 
 
& Possible Typo & 2.350 (8.098) & 0.041 (0.102) \\ 
 
& Grammar Errors & 0.195 (0.541) & 0.004 (0.012) \\ 
 
& Punctuation Errors & 0.654 (2.096) & 0.010 (0.048) \\ 
 
& Typographic Errors & 0.758 (2.356) & 0.015 (0.071) \\ 
 
& Style Errors & 0.343 (0.933) & 0.004 (0.012) \\ 
 
& Miscellaneous Errors & 0.094 (0.353) & 0.002 (0.008) \\ 
 
& Redundant Phrases & 0.027 (0.172) & 0.000 (0.003) \\ 
 
& Nonstandard Phrases & 0.002 (0.052) & 0.000 (0.001) \\ 
 
& Commonly Confused Words & 0.008 (0.093) & 0.000 (0.002) \\ 
 
& Collocations & 0.013 (0.125) & 0.000 (0.003) \\ 
 
& Semantic Errors & 0.007 (0.113) & 0.000 (0.001) \\ 
 
\

%% file: tables/predict_hiring_languagetool.tex
\begin{table}[!htbp] \centering 
  \caption{Hiring outcomes predicted based on language errors in the control group } 
  \label{tab:predict_hiring_languagetool} 
\small 
\begin{tabular}{@{\extracolsep{5pt}}lcc} 
\\[-1.8ex]\hline 
\hline \\[-1.8ex] 
 & \multicolumn{2}{c}{\textit{Dependent variable:}} \\ 
\cline{2-3} 
\\[-1.8ex] & Number of Contracts & Hired \\ 
\\[-1.8ex] & (1) & (2)\\ 
\hline \\[-1.8ex] 
 Capitalization Error & $-$0.010$^{***}$ & $-$0.006$^{***}$ \\ 
  & (0.002) & (0.001) \\ 
  Possible Typo & 0.0002 & 0.0001 \\ 
  & (0.0001) & (0.0001) \\ 
  Grammar Error & $-$0.002 & $-$0.001 \\ 
  & (0.002) & (0.001) \\ 
  Punctuation Error & 0.006$^{***}$ & 0.003$^{***}$ \\ 
  & (0.001) & (0.0003) \\ 
  Typography Error & $-$0.001 & $-$0.001$^{**}$ \\ 
  & (0.0005) & (0.0002) \\ 
  Style Error & 0.010$^{***}$ & 0.006$^{***}$ \\ 
  & (0.001) & (0.001) \\ 
  Miscellaneous Error & $-$0.010$^{***}$ & $-$0.004$^{***}$ \\ 
  & (0.003) & (0.002) \\ 
  Redundant Phrases & 0.019$^{***}$ & 0.012$^{***}$ \\ 
  & (0.007) & (0.003) \\ 
  Nonstandard Phrases & 0.071$^{***}$ & 0.026$^{**}$ \\ 
  & (0.022) & (0.011) \\ 
  Commonly Confused Words & $-$0.028$^{**}$ & $-$0.014$^{**}$ \\ 
  & (0.012) & (0.006) \\ 
  Collocations & $-$0.007 & $-$0.006 \\ 
  & (0.009) & (0.005) \\ 
  Semantic Error & $-$0.014 & $-$0.007 \\ 
  & (0.010) & (0.005) \\ 
  Constant & 0.140 & 0.152$^{**}$ \\ 
  & (0.142) & (0.072) \\ 
 \hline \\[-1.8ex] 
Controls & X & X \\ 
Observations & 93,725 & 93,725 \\ 
R$^{2}$ & 0.004 & 0.005 \\ 
\hline 
\hline \\[-1.8ex] 
\end{tabular}
\\
\begin{minipage}{\textwidth}
{\footnotesize \emph{Notes}: This table analyzes correlation between various writing errors on jobseekers' resumes and their hiring outcomes. 
          Column (1) defines Number of Contracts as the number of unique jobs they work over the month after they register for the platform.
          Column (2) defines Hired as the probability the jobseekers was hired over that month. Column (3) is the number of interviews they gave over that month. 
          And the Column (4) outcome Invitations is the number of times they were recruited to a job over their first month.
          All analysis includes controls for profile hourly rate and job category. Writing errors are defined by LanguageToolR. 
          The sample is made up of all jobseekers in the control group of the experimental sample who submitted non-empty resumes and were approved by the platform.  \newline \starlanguage}
\end{minipage}
\end{table}

%% file: tables/predict_hiring_languagetool_total.tex
\begin{table}[!htbp] \centering 
  \caption{Hiring outcomes predicted based on language errors in the control group } 
  \label{tab:predict_hiring_languagetool_total} 
\small 
\begin{tabular}{@{\extracolsep{5pt}}lcccccccc} 
\\[-1.8ex]\hline 
\hline \\[-1.8ex] 
 & \multicolumn{8}{c}{\textit{Dependent variable:}} \\ 
\cline{2-9} 
\\[-1.8ex] & \multicolumn{2}{c}{Number of Contracts} & \multicolumn{2}{c}{Hired} & \multicolumn{2}{c}{Interviews} & \multicolumn{2}{c}{Invitations} \\ 
\\[-1.8ex] & (1) & (2) & (3) & (4) & (5) & (6) & (7) & (8)\\ 
\hline \\[-1.8ex] 
 Total errors & 0.001$^{***}$ &  & 0.0003$^{***}$ &  & 0.003$^{***}$ &  & 0.002$^{***}$ &  \\ 
  & (0.0001) &  & (0.0001) &  & (0.0003) &  & (0.0003) &  \\ 
  Error rate &  & $-$0.048$^{***}$ &  & $-$0.031$^{***}$ &  & $-$0.162$^{***}$ &  & $-$0.103$^{***}$ \\ 
  &  & (0.008) &  & (0.004) &  & (0.021) &  & (0.017) \\ 
  Constant & 0.162 & 0.171 & 0.164$^{**}$ & 0.169$^{**}$ & $-$0.025 & 0.013 & $-$0.015 & 0.009 \\ 
  & (0.142) & (0.142) & (0.072) & (0.072) & (0.380) & (0.380) & (0.305) & (0.305) \\ 
 \hline \\[-1.8ex] 
Normalized & N & Y & N & Y & N & Y & N & Y \\ 
Controls & X & X & X & X & X & X & X & X \\ 
Observations & 93,725 & 93,725 & 93,725 & 93,725 & 93,725 & 93,725 & 93,725 & 93,725 \\ 
R$^{2}$ & 0.002 & 0.002 & 0.002 & 0.003 & 0.006 & 0.006 & 0.004 & 0.004 \\ 
\hline 
\hline \\[-1.8ex] 
\end{tabular}
\\
\begin{minipage}{\textwidth}
{\footnotesize \emph{Notes}: This table analyzes correlation between all writing errors on jobseekers' resumes and their hiring outcomes. 
          The first independent variable is the total number of writing errors on a jobseekers' resume. The second independent variable is the total number of errors divided by the length of their resume, in number of words.
          Columns (1) and (2) define Number of Contracts as the number of unique jobs they work over the month after they register for the platform.
          Columns (3) and (4) defines Hired as the probability they were hired over that month. Columns (5) and (6) is the number of interviews they gave over that month. 
          And the Columns (7) and (8) outcome Invitations is the number of times they were recruited to a job over their first month.
          All analysis includes controls for profile hourly rate and job category. Writing errors are defined by LanguageToolR. 
          The sample is made up of all jobseekers in the control group of the experimental sample who submitted non-empty resumes and were approved by the platform.  \newline \starlanguage}
\end{minipage}
\end{table}

%% file: tables/writing_outcomes_1.tex
\begin{table}[!htbp] \centering 
  \caption{Effect of Treatment to Writing Errors, Page 1} 
  \label{tab:writing_outcomes_1} 
\small 
\begin{tabular}{@{\extracolsep{5pt}}lccccccc} 
\\[-1.8ex]\hline 
\hline \\[-1.8ex] 
 & \multicolumn{7}{c}{\textit{Dependent variable:}} \\ 
\cline{2-8} 
\\[-1.8ex] & Spelling & Capitalization & Possible Typo & Grammar & Punctuation & Typography & Style \\ 
\\[-1.8ex] & (1) & (2) & (3) & (4) & (5) & (6) & (7)\\ 
\hline \\[-1.8ex] 
 Algo Writing Treatment & $-$0.001$^{*}$ & $-$0.0005$^{***}$ & $-$0.002$^{***}$ & $-$0.0005$^{***}$ & $-$0.0004 & $-$0.002$^{***}$ & 0.0003$^{***}$ \\ 
  & (0.0005) & (0.0001) & (0.0005) & (0.0001) & (0.0004) & (0.0003) & (0.0001) \\ 
  Constant & 0.036$^{***}$ & 0.003$^{***}$ & 0.041$^{***}$ & 0.004$^{***}$ & 0.010$^{***}$ & 0.015$^{***}$ & 0.004$^{***}$ \\ 
  & (0.0003) & (0.00005) & (0.0003) & (0.00004) & (0.0003) & (0.0002) & (0.00004) \\ 
 \hline \\[-1.8ex] 
Observations & 187,857 & 187,857 & 187,857 & 187,857 & 187,857 & 187,857 & 187,857 \\ 
R$^{2}$ & 0.00002 & 0.0003 & 0.0001 & 0.0004 & 0.00000 & 0.0004 & 0.0001 \\ 
\hline 
\hline \\[-1.8ex] 
\end{tabular}
\\
\begin{minipage}{ \textwidth}
{\footnotesize \emph{Notes}: 
          This table analyzes the effect of the treatment on all types of writing errors, normalized by resume length. Writing errors are defined by LanguageToolR, and divided by the number of words in a jobseekers' resume to calculate their error rate. 
          The sample is made up of all jobseekers in the experimental sample who completed the platform registration page and submitted non-empty resume.
          \starlanguage}
\end{minipage}
\end{table}

%% file: tables/writing_outcomes_2.tex
\begin{table}[!htbp] \centering 
  \caption{Effect of Treatment to Writing Errors, Page 2} 
  \label{tab:writing_outcomes_2} 
\small 
\begin{tabular}{@{\extracolsep{5pt}}lcccccc} 
\\[-1.8ex]\hline 
\hline \\[-1.8ex] 
 & \multicolumn{6}{c}{\textit{Dependent variable:}} \\ 
\cline{2-7} 
\\[-1.8ex] & Miscellaneous & Redundant Phrases & Nonstandard Phrases & Commonly Confused Words & Collocations & Semantics \\ 
\\[-1.8ex] & (1) & (2) & (3) & (4) & (5) & (6)\\ 
\hline \\[-1.8ex] 
 Algo Writing Treatment & $-$0.0004$^{***}$ & $-$0.00003$^{**}$ & $-$0.00001 & $-$0.00005$^{***}$ & $-$0.0001$^{***}$ & 0.00001 \\ 
  & (0.00003) & (0.00001) & (0.00000) & (0.00001) & (0.00001) & (0.00001) \\ 
  Constant & 0.002$^{***}$ & 0.0004$^{***}$ & 0.00003$^{***}$ & 0.0001$^{***}$ & 0.0003$^{***}$ & 0.0001$^{***}$ \\ 
  & (0.00002) & (0.00001) & (0.00000) & (0.00001) & (0.00001) & (0.00000) \\ 
 \hline \\[-1.8ex] 
Observations & 187,857 & 187,857 & 187,857 & 187,857 & 187,857 & 187,857 \\ 
R$^{2}$ & 0.001 & 0.00002 & 0.00001 & 0.0002 & 0.0001 & 0.00000 \\ 
\hline 
\hline \\[-1.8ex] 
\end{tabular}
\\
\begin{minipage}{ \textwidth}
{\footnotesize \emph{Notes}: 
          This table analyzes the effect of the treatment on all types of writing errors, normalized by resume length. Writing errors are defined by LanguageToolR, and divided by the number of words in a jobseekers' resume to calculate their error rate. 
          The sample is made up of all jobseekers in the experimental sample who completed the platform registration page and submitted non-empty resume.
         \starlanguageexpanded} 
\end{minipage}
\end{table}